\documentclass[fleqn,usenatbib]{mnras}
\usepackage{header_mnras}
\usepackage{affiliations}
\newcommand{\revone}[1]{#1}

\title[HII regions in NGC~2366]{Chemical abundances and small-scale gas kinematics in the local star-forming galaxy NGC~2366}

\author[Yarovova et al.]{
Anastasiya D. Yarovova$^{1, 2}$\hyperref[fn:email]{\thanks{E-mail: \href{mailto:yaan.ph@gmail.com}{yaan.ph@gmail.com} (AY); \href{mailto:ivan.gerasimov@oca.eu}{ivan.gerasimov@oca.eu} (IG); \href{mailto:oleg.egorov@uni-heidelberg.de}{oleg.egorov@uni-heidelberg.de} (OE).}}\label{fn:email}\hyperref[fn:equal]{\thanks{These authors contributed equally to this work.}}\label{fn:equal}%\thanks{Email: \href{mailto:yaan.ph@gmail.com}{yaan.ph@gmail.com}},
Ivan~S.~Gerasimov$^{3, 1}$\hyperref[fn:email]{\footnotemark[1]}\hyperref[fn:equal]{\footnotemark[2]}%\thanks{Email: \href{mailto:ivan.gerasimov@oca.eu}{ivan.gerasimov@oca.eu}}, 
Oleg V. Egorov$^{4}$\hyperref[fn:email]{\footnotemark[1]}%\thanks{Email: \href{mailto:oleg.egorov@uni-heidelberg.de}{oleg.egorov@uni-heidelberg.de}}, 
Alexei~V.~Moiseev$^{2, 1}$, \newauthor
Konstantin~I.~Vasiliev$^{1}$, 
Tatiana~A.~Lozinskaya$^{1}$\\
$^1$ \SAI \\
$^2$ \SAO \\ 
$^3$ \Nice \\ 
$^4$ \HD \\
}
\date{Accepted XXX. Received YYY; in original form ZZZ}

\pubyear{2025}

\begin{document} 
\label{firstpage}

% Uncomment for MNRAS, comment for AA
\pagerange{\pageref{firstpage}--\pageref{lastpage}}
\maketitle

% Abstract of the paper (use \abstract for aa, \begin{abstract} for MNRAS
\begin{abstract}
We present a detailed spectroscopic study of ionised gas in the nearby ($D\sim3.3$Mpc) dwarf galaxy NGC~2366, a local analogue of Green Pea galaxies, based on  observations with the SCORPIO-2 instrument at the Russian 6-m BTA telescope. Using scanning Fabry-Perot interferometry and long-slit spectroscopy, we examine the gas kinematics, excitation mechanisms, and chemical abundances across the disc of NGC~2366, 
including its prominent starburst region Mrk~71 and the companion region NGC~2363. %, with the spatial resolution $\sim30$ pc. 
We identified 20 regions with locally elevated \Ha{} velocity dispersion, only four of which correspond to known high-energy sources. We argue that one of the remaining objects can be a previously unidentified Wolf--Rayet star and two -- supernova remnants. For 15 \HII{} regions, we measure electron temperatures, oxygen and nitrogen abundances via the `direct' $T_e$ method, with $12 + \log(\mathrm{O}/\mathrm{H})$ ranging from 7.6 to 8.0 in most of the regions. We show that Mrk~71 has higher oxygen abundance compared to the other \HII{} regions in the galaxy, contrary to the previous indirect estimates \revone{suggesting flat gradient throughout the galaxy}. Together with the localized spatial variations of metallicity in the area, it is indicative of metal enrichment by the outflow from the super star cluster in the centre of Mrk~71. % which could point to a potential enrichment of the galactic ISM by metals through the outflow in Mrk 71. Detection of chemical abundance variations throughout the galactic disk contradicts previous indirect measurements. %We identify a new strong candidate Wolf--Rayet star in the northern galaxy outskirts, two new SNR candidates and one X-ray binary star candidate.
%Our results suggest spatially inhomogeneous chemical enrichment and widespread mechanical feedback in NGC~2366.
\end{abstract}

% uncomment for aa; comment for mnras
% \keywords{
% galaxies: abundances -- galaxies: individual: NGC~2366  }
%\maketitle

% uncomment for mnras; comment for aa
\begin{keywords}
galaxies: abundances --- galaxies: individual: NGC~2366 --- stars: Wolf–Rayet --- ISM: supernova remnants

\end{keywords}

\section{Introduction}\label{introduction}

\begin{figure*}
    \centering
    \includegraphics[width=\linewidth]{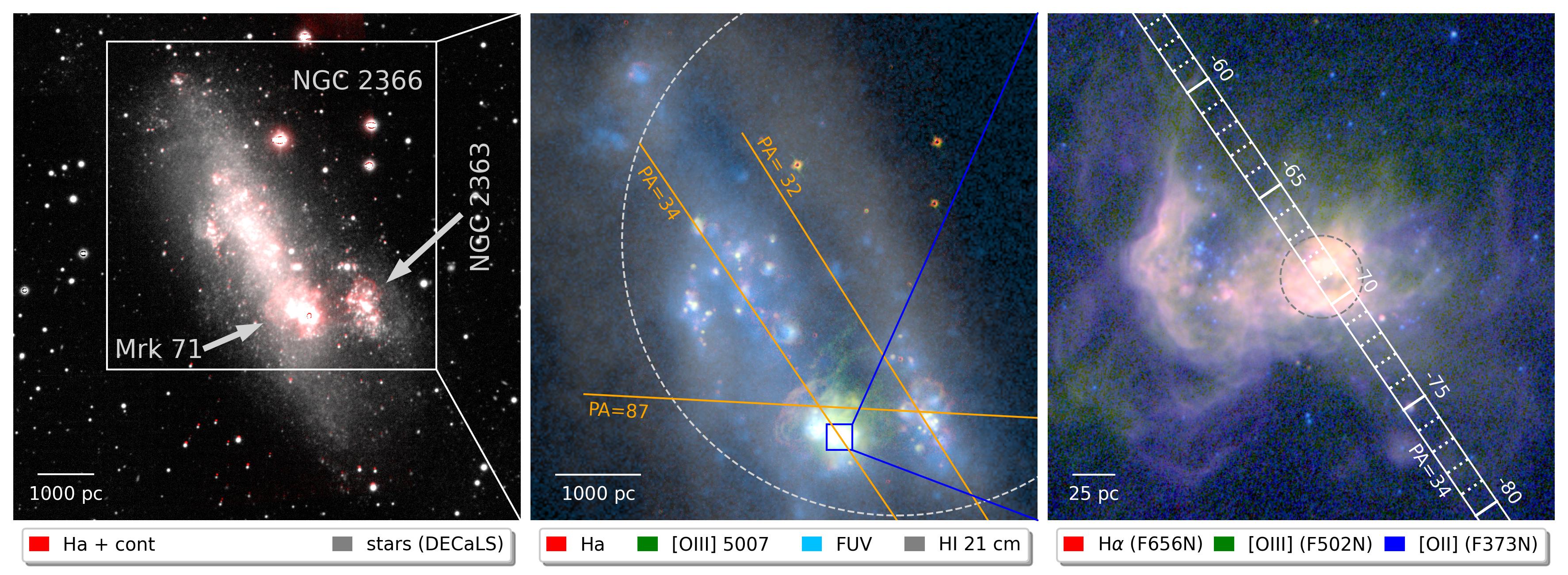}
    \caption{The image of NGC~2366 (left-hand panel) and zoom-in to its central part (middle panel) and Mrk~71 (right-hand panel). Left panel shows \Ha{}+continuum narrow-band image (red; this paper) combined with the stellar continuum (grey; DECaLS; \citealt{Dey2019}). Middle panel shows combination of images in \Ha{} (red; this paper) and  \OIII{} (green; this paper) emission lines tracing ionized gas, far-UV (blue; GALEX; \citealt{Hunter2010}) tracing young stellar population, and \HI{} 21~cm (grey;  LITTLE THINGS; \citealt{Hunter2012}) tracing atomic gas. Right-hand panel shows 3-color image of the Mrk~71 region obtained from HST narrow-band images F656N (centered at \Ha{}; red), F502N (centered at \OIII{}; green) and F373N (centered at \OII{}; blue) filters (from \citealt{James2016}). Localization of the SCORPIO-2 long slit positions are shown by yellow lines in the middle panel, with zoom-in on the Mrk~71 region for the slit PA=34$\degr$ in the left-hand panel. White dashed line in the middle panel shows the field of view for FPI data. Grey circle in the right-hand panel indicates the shell around knot A \citep{James2016}.
    % discussed in the text (see Sec.~\ref{sec:analisys_spec}).
    }\label{fig:morphology}
\end{figure*}

% Dwarf galaxies are excellent objects for studying the processes associated with massive stars, such as producing of UV radiation, chemical enrichment of ISM and supernova explosions. These processes influence the evolution of the host galaxy as well as the intergalactic medium. Occurring under conditions of reduced metallicity, they have the least certainty, therefore being of significant interest.

%Feedback impact from massive stars onto the interstellar medium (ISM) plays a crucial role during the galaxy evolution and the most importantly regulates star formation during the epoch of reionisation. Study of mechanical contribution of massive stars and their impact on metal enrichment can put constrains on the key parameters regulating interaction between different components of a galaxy.  

During their evolution, massive stars inject energy, momentum, and products of their nucleosynthesis into the interstellar medium (ISM) via stellar winds, radiation, cosmic rays, and supernova explosions, and thus play a crucial, although still not totally understood, role in the regulation of gas morphology, dynamics, star formation, metal enrichment, and overall galaxy evolution \citep[e.g., see reviews by][]{Krumholz2014, Schinnerer2024}. 
These effects can be studied in nearby dwarf irregular (dIrr) galaxies that not only provide us access to a metal-poor environment, but also have other advantages such as gas-rich discs and the absence of spiral density waves washing out smaller-scale structures \citep[e.g., see review by][]{Hunter2024}. Among dwarf galaxies there is a subclass of Green Pea galaxies --  starbursts at redshifts $\rm{z}\sim0.2$. They exhibit exceptionally strong leakage of Lyman continuum photons, which are considered 
%are also those that demonstrate strong outflows and share a number of parameters with 
to be responsible for the reionization of the early Universe \citep{Ouchi2009, Martin2024}. %One of such galaxies we aim to study in the present work.

In this work, we study ionized gas and massive stars in the nearby dwarf galaxy NGC~2366, which is known to be the nearest analogue of Green Pea galaxies \citep{Micheva2017}. It belongs to the M~81 group and is located at a distance of 3.34~Mpc \citep{Tully2013}. The proximity of the galaxy (the spatial scale is 16.1 pc per arcsec) enables resolved studies of HII regions and individual massive stars.

Over the past 1~Gyr the galaxy shows high star formation (SF) activity ($\sim 0.03~\rm M_\odot$/yr) with two peaks up to $\rm 0.15~M_\odot$/yr. One was 500~Myr ago and the second is the ongoing starburst \cite[see][]{Weisz2008, McQuinn2010}. \cite{Feinstein2024} identified 23 young ($< 10$~Myr) star clusters in the galaxy, confirming that ongoing star formation is quite active. %Stellar population studies for NGC~2366 were also carried out by several groups (study of bright stars: \citealt{Aparicio1995,Troshynski2024} and references therein; presence of variable stars \citealt{Davidge2022,Tolstoy1995}).
Individual stars in NGC~2366 were extensively studied \citep[e.g.][]{Aparicio1995,Tolstoy1995, Davidge2022, Troshynski2024} including population of young massive WR, LBV and Very Massive Stars (VMS) \citep[e.g.][]{Drissen1997, Drissen2000, Smith2023, Oey2023}. The galaxy hosts three supernova remnants found both from optical \citep{Vucetic2019, Ercan2021} and radio-continuum \citep{Chomiuk2009} observations. %(two candidates listed in \citealt{Vucetic2019}, four candidates in \citealt{Ercan2021}) and in radio (five candidates detected by \citealt{Chomiuk2009}). 

In NGC~2366 particular attention was drawn to its brightest region, responsible for ionization and shaping of its ISM -- Mrk~71 (e.g. \citealt{James2016, Micheva2017, Oey2017, Micheva2019, Komarova2021, Chen2023, MendezDeldago2024}). Together with 30~Dor and NGC~5253, it is one of the iconic starburst regions in the Local Universe. Mrk~71 consists of super star clusters knot~A and knot~B (SSCs, young compact massive clusters with M $\sim\mathrm{10^5~M_\odot}$ that can be progenitors of the globular clusters). Knot~A potentially hosts VMS (\citealt{Smith2023}, \citealt{Oey2023}), while knot~B contains the population of WR stars \citep{Drissen2000}, and one LBV close to it \citep{Drissen1997, Drissen2001}. \cite{Thuan2014} and \cite {Kaaret2024} observed NGC~2366 in X-ray, which resulted in a detection of emission from a point source located %in the tip of outflow 
in the knot B, which could be potentially connected to the accretion onto a compact object. 
%\cite{Oey2017} report the detection of 5 CO clouds with the Northern Extended Millimeter Array. One of the clouds coincides with Mrk~71-A super star cluster (SSC), having mass of $\mathrm{10^5M_\odot}$, while four others are located within 50 pc of it. 

% Mrk~71 was studied in detail with many multi-wavelength observational facilities, including integral-field spectroscopy with Keck/KCWI \citep{Chen2023}, extensive optical imaging \citep{James2016} and ultraviolet (UV) spectral observations \citep{Drissen2000,Smith2023, Oey2023} with HST. These studies resulted in finding of characteristic features of the object.

Studies of small-scale kinematics in NGC~2366 of ionized gas have mainly focused on Mrk~71. \cite{Roy1991} analyzed \OIII{}$\lambda5007$\AA\ line profile using a scanning Fabry-Perot interferometer (FPI), and detected ionized gas outflow from Mrk~71 region exhibiting underlying broad-line component in \OIII{} line. %Later, kinematical studies of the Mrk~71 region were continued by \cite{Roy1992}, \cite{GonzalezDelgado1994}, \cite{Yang1994}.
This outflow was confirmed in follow-up studies \citep{Roy1992, GonzalezDelgado1994, Yang1994}. Using a deep \Ha{} image, \cite{vanEymeren2007} showed that the outflow has conical form and is extended up to $\sim 900$~pc from knot B. % more extended, than assumed before ($\sim 900$~pc).
\cite{Micheva2019} performed integral-field spectroscopy of Mrk~71 with fiber-fed Potsdam Multi-Aperture Spectrophotometer (PMAS) and estimated the maximum FWHM in the \Ha{} line to be equal to $1100 \kms$ and also suggested that the related outflow could possibly be biconical. 
As an explanation of the broad component, \cite{Binette2009} and \cite{Komarova2021} 
 proposed turbulent mixing layers and radiation-driven superwind from Knot A, respectively.

Another bright star-forming region -- NGC~2363\footnote{note that NGC~2363 is often mistakenly called as Mrk~71 region in the literature, while correctly it refers to the extended star-forming region in the South-East part of NGC~2366 galaxy (see \citealt{Micheva2019} and Fig.~\ref{fig:morphology})} -- is sometimes considered to be part of the satellite galaxy, whose passage could initiate the starburst in NGC~2366 \citep{Drissen2000}. However, \HI{} observations do not show significant perturbations in the atomic gas velocity field that would suggest an interaction \citep{Hunter2001}, leaving open the possibility that it is a star-forming complex in the inclined NGC 2366 disc.  %These authors also point out that the kinematical analysis shows the disconnection of NGC~2363 from the main body of NGC~2366. However, \cite{vanEymeren2009} showed that \HI{} kinematics can be described by a simple rotation model (their fig.~6) and it is consistent with the ionized gas velocity traced by \Ha{} (their fig.~7).

The oxygen abundance (a proxy of gas-phase metallicity) of Mrk~71 was previously measured to be $\mathrm{12 + \log(O/H) \sim 7.8 - 7.9}$   \citep[e.g.][]{GonzalezDelgado1994, Noeske2000, Esteban2009, Izotov2011, Chen2023}. Based on the narrow-band HST images, \citet{James2016} %\cite{James2016} obtained narrow-band images in \OII, \HeII, \Hb, \OIII, \Ha, and \SII{} of the Hubble Space Telescope Wide Field Camera 
 revealed chemical inhomogeneity in Mrk~71 on scales of less than 50 pc. Recent KCWI observations, however, revealed no evidence of chemical or temperature inhomogeneities \citep{Chen2023}, though this finding was later challenged by \citet{MendezDeldago2024}. Meanwhile, the rest of the galaxy has yet to be studied through spectroscopic observations: the only indirect estimates of gas-phase metallicity in the NGC~2366 disc were provided by \citet{Roy1996} based on narrow-band photometrical data. Their estimate of $\mathrm{12 + \log (O/H) \sim 8.2}$ suggests that Mrk~71 is significantly more metal-poor than the rest of the NGC~2366 disc.
 %Such measurements could shed light on the interactions of Mrk~71 with the surrounding gas as well as provide clues on metal enrichment from stellar feedback driven outflows. In addition, small-scale (20--50 pc) kinematics across the entire galaxy field %, which is of particular interest in connection with
% can reveal the numerous high-energy source candidates.%, have not been studied yet. 

% (\cite{Thuan2005}, \cite{McQuinn2010}).
% Stellar studies: \cite{Troshynski2024} (massive stars, photometry)
% 15. FUV spectra, characteristic features of very massive stars \cite{Smith2023}
% \cite{Davidge2022} red variable stars
% \cite{Tolstoy1995} variable stars
% \cite{Aparicio1995} The stellar content of the Im galaxy NGC~2366 is discussed on the basis of CCD BVR photometry. 
%%%%%%%%%%%%%%%%%%%%%%%%%%%%%%%%%%%%%%%%%%%%%%%
%%%%%%%%%%%%%%%%%%%%%%%%%%%%%%%%%%%%%%%%%%%%%%%
%%%%%%%%%%%%%%%%%%%%%%%%%%%%%%%%%%%%%%%%%%%%%%%

\begin{table*}
\small
    \centering
	\caption{ Log of observations: name of data set, date of observations, total exposure time $T_{exp}$, field of view (FOV), pixel scale, angular resolution ($\theta$),   spectral range ($\Delta\lambda$) and  spectral resolution ($FWHM$) }
	\label{obs:log}	
    \begin{tabular}{c|c|c|c|c|c|c|c}
	\hline
	Data set              & Date     & $T_{exp}$, s & FOV                             & pixel scale, $''$             & $\theta$, $''$   & $\Delta\lambda$ or $\lambda_c$, \AA                 & $\mathrm{FWHM}$, \AA     \\ 
\hline
                                  
    \multicolumn{8}{c}{Narrow-band imaging} \\    
    CMO \OIII\ 5007~\AA & 2018 Feb 15,17  & 2100 & $11.1\times10.6$  & 0.155 &  0.9--1.2     &  4992  & 120.8 \\
   CMO \OIII\ continuum & 2018 Feb 15,17  & 1500 & $11.1\times10.6$  & 0.155 &  0.9--1.2     &  5117.8  & 94.6 \\
   CMO \Ha+\NII & 2018 Feb 15,17,21  & 1800 & $11.1\times10.6$  & 0.155 &  0.9--1.2     &  6559  & 76.5 \\
   CMO \Ha\ continuum & 2018 Feb 15  & 600 & $11.1\times10.6$  & 0.155 &  0.8--1.0     &  6426.9  & 122.6 \\
   CMO \SII\ 6717+6731 \AA & 2018 Feb 21  & 2400 & $11.1\times10.6$  & 0.155 &  1.0-1.25     &  6721.3  & 56.8 \\
   CMO \SII\ continuum & 2018 Feb 21  & 900 & $11.1\times10.6$  & 0.155 &  1.0-1.25     &  6959.7  & 153 \\
 
  \multicolumn{8}{c}{\revone{LS} spectroscopy}\\
	SCORPIO-2  PA=$34\degr$ & 2017 Dec 10  & 2100  & {$1\arcsec\times6.1\arcmin$} & 0.36 &  2.8  &  3650-7200 & 4.5  \\
	SCORPIO-2  PA=$32\degr$ & 2017 Dec 10  & 3600 & {$1\arcsec\times6.1\arcmin$} & 0.36 & 2.0   &  3650-7200 & 4.5 \\
	SCORPIO-2  PA=$87\degr$ & 2023 Mar 18 & 6300  & {$1\arcsec\times6.8\arcmin$} & 0.4  &  2.8  & 3650-7200  & 4.5  \\
        \multicolumn{8}{c}{FPI data} \\    

    SCOPRIO-2 FPI &  2022 Dec 31 & 150 $\times$ 40 & $6.8\arcmin\times6.8\arcmin$ & $0.4\arcsec$ &  $1.8\arcsec$     & 8.8 \AA{} around \Ha{}  & 0.48 \\
    \hline
	\end{tabular}
\end{table*}

We aim to deeper study of the ionized gas and its connection to massive stars beyond the Mrk~71 area. In this study, we investigate the local gas kinematics, excitation mechanisms, and gas-phase metallicity of the star-forming regions across the entire disc of NGC~2366, including Mrk~71 and NGC~2363 regions. We present here the new spectral observations obtained with three long-slit \revone{(LS)} positions and high spectral resolution Fabry-Perot interferometer in the \Ha{} emission line, as well as the photometric observations narrow-band filters centred on key emission lines. The goal of these observations was to determine the nature of the objects with local elevation of the \Ha{} velocity dispersion and to measure their physical properties and chemical abundances ($\mathrm{T_e,~ n_e,~ 12+log(O/H), ~log(N/O)}$). %Our long-slit spectral data covers Mrk~71 and the related outflow region, as well as the NGC~2363 star-forming region and the kinematically peculiar object in the northern part of the galaxy, so we can compare different regions using homogeneous data and uniform methods. 

The paper is organized as follows: in Section~\ref{sec:obs_and_dat} we provide details on the observations and data reduction; Section~\ref{sec:analysis} describes the analysis of the data: identification of \HII{} regions, analysis of the local gas kinematics and ionization state of emission nebulae, calculation of the gas-phase metallicity; in Section~\ref{sec:discussion}, we discuss the obtained results and draw conclusions on the nature of the studied \HII{} regions; in Section~\ref{sec:summary}, we summarize our results.
% 12. https://ui.adsabs.harvard.edu/abs/2016ApJ...816...40J/abstract наблюдения с HST-WFC3
% 14. \cite{Kaaret2024}: "We observed NGC~2366 with the Chandra X-ray Observatory and detected X-ray emission from a point source coincident with the apex of the outflow at knot B. The pointlike nature and variability of the X-ray emission suggests accretion onto a compact object. The accretion could produce sufficient kinetic energy to power the outflow."
% 15. Недавнее про Knot A и Knot B \cite{Danehkar2023}
% 16. Здесь тоже спектры через Mrk~71 \cite{Hunter1997}

\section{Observations and data reduction}
\label{sec:obs_and_dat}

\subsection{3D spectroscopy with a scanning Fabry-Perot Interferometer}
\label{subsec:FPI}

Observations with scanning Fabry-Perot interferometer (FPI) were performed with SCORPIO-2 multi-mode focal reducer \citep{scorpio2} installed in the prime focus of the BTA 6-m telescope of Special Astrophysical Observatory of Russian Academy of Science (SAO RAS). During the observations, we consecutively obtained 40 interferograms with the different gap between the FPI plates. The narrow-band filter with FWHM $\sim 14$~\AA\ was used to cut the operating spectral range around the \Ha{} line. Details of the observations are given in Table~\ref{obs:log}. 

Data reduction was performed with the pipeline in \textit{IDL} environment described in a series of papers \citep{Moiseev2002, Moiseev2008, Moiseev2021}. After primary reduction, sky subtraction, photometric and seeing corrections, the data were combined into a single 40-channel cube with angular sampling of $0.4\arcsec$ and spectral sampling of $0.22$~\AA. The resulting seeing-limited angular resolution is $\sim1.8$~ arcsec, and the spectral resolution is $\sim0.48$~\AA\ ($\sim22\kms$).

Following \citealt{Moiseev2008}, we fitted a single-component  Voigt function to the data cube to obtain 2D distributions of the \Ha\ flux, velocity and intrinsic velocity dispersion. %corrected for instrumental broadening. 
The multi-component Voigt profile was fitted to the data in the regions exhibiting  prominent asymmetry of the line \Ha{} if the residuals after subtracting the results of a single-component fit exceeded $3\sigma$. 
% In several cases, where our analysis revealed asymmetry in line profiles, we used fitted a multi-component Voigt profile to the data cubes in these regions. Note, that we try the least number of components on the $3\sigma$ error basis.

\subsection{Long-slit spectroscopy}
\label{subssec:longslit}

\revone{LS} spectroscopy was performed with the same SCORPIO-2 focal reducer at the SAO RAS  6-m telescope. Details of the observations are given in Table  \ref{obs:log}. Three spectra with positional angles PA=32$\degr$, 34$\degr$ and 87$\degr$ were obtained in different epochs with the same instrumental setup: 1 arcsec slit width, the spectral range 3650-7200 \AA{} with a resolution of 4.5 \AA. Localization of the slits is shown in Fig.~\ref{fig:morphology}.% We used VPHG1200@540 grating and a slit of 1 arcsec width. The data obtained cover the spectral range 3650-7200 \AA{} with a spectral resolution of 4.5 \AA.  

Data reduction was performed in a standard way using an IDL-based pipeline for SCORPIO data. The procedure performed is described in the previous works \citep[for example][]{Egorov2018}. For absolute flux calibration, we used the spectra of spectrophotometric standard stars (G191B2B and BD+75D325) obtained during the same nights and located at a close zenith distance right before the object observations. 

%For further processing we used our own scripts running on {\sc python}. 
% We extracted the spectra of all regions crossed by long-slits by integrating in the apertures along the slit, corresponding to 2-4 seeing values. The integral fluxes in the emission lines were measured by fitting a single-component Gaussian into emission line profiles. The $1\sigma$ uncertainties of the measured fluxes are estimated with repeating measurements for 100 Monte-Carlo simulated spectra. The measured fluxes are corrected for reddening using the \textsc{pyneb} {\sc python} package \citep{Luridiana2015} assuming \citet{Fitzpatrick1999} reddening curve. The $E({B-V})$ value is estimated from the Balmer decrement with the coefficient \Ha/\Hb{} consistent with the temperature estimations obtained by the $T_e$ method. We report the measured fluxes and the corresponding errors for each region analyzed in this paper in Tables ~\ref{table_fluxes1}, \ref{table_fluxes2}, \ref{table_fluxes_custom}.

\subsection{Narrow-band imaging}
\label{subsec:narrowband}

\begin{figure*}
\includegraphics[width=\linewidth,trim={0 0 0.65cm 0},clip]{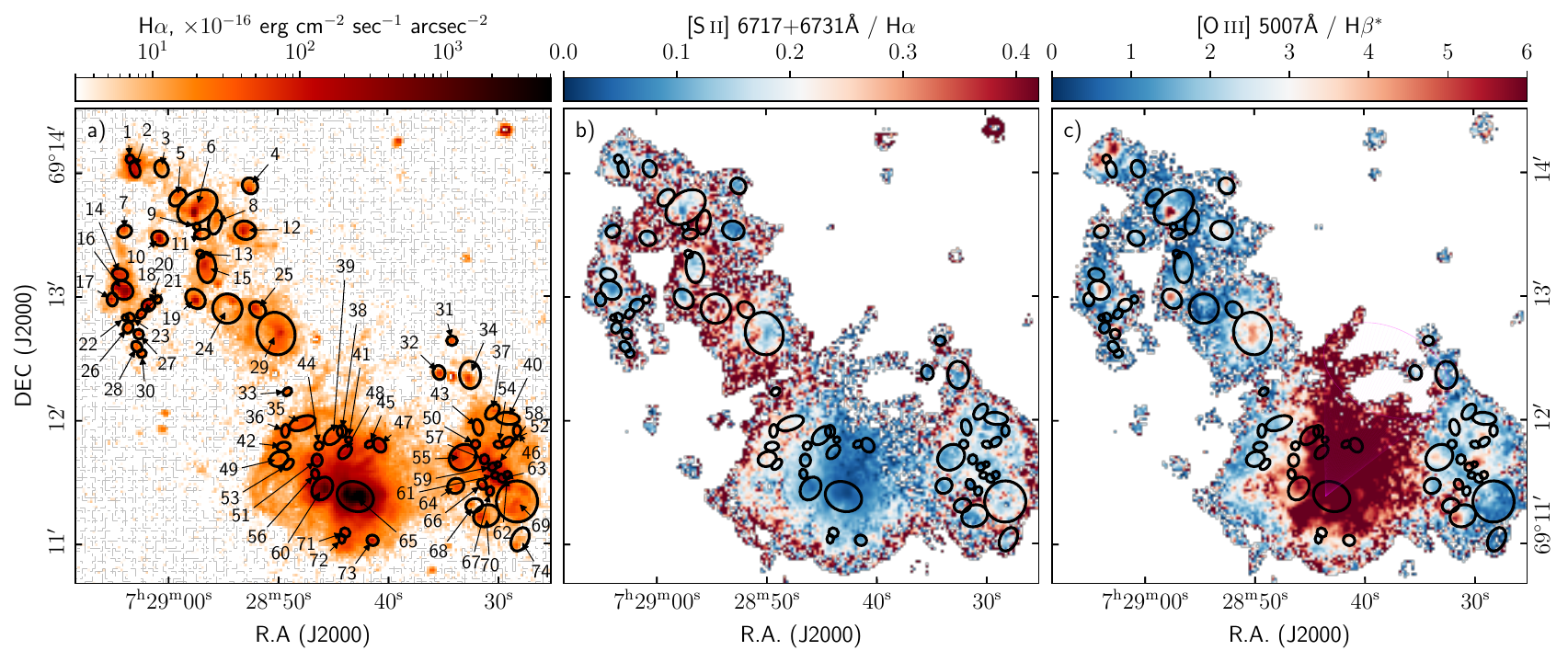}
\caption{Maps of NGC~2366 obtained from narrow-band images in \Ha{}, \SII{} and \OIII{} emission lines. Panel `a' shows a map of NGC~2366 in \Ha{} line with overlapped \HII{} region positions defined in this work. Panels `b' and `c' show \SII{}/\Ha{} and \OIII{}/\Hb{} ratios, respectively. \Hb{}* is obtained as \revone{\Ha{}/3.1 (which is consistent with median \Ha/\Hb, obtained from spectra)}. Mrk~71 and its outflow region (shown by magenta lines on panel `c') are clearly outstanding showing high degree of ionization (high \OIII{}/\Hb{}) and low \SIIHa{} ratio.
}
\label{fig:lineratiomaps}
\end{figure*}

We analyze here the images obtained at the 2.5-m  telescope of the Caucasian Mountain Observatory (CMO) in \OIII{}, \SII{} and \Ha{} emission lines 
 and the adjacent stellar continuum with the 
Wide Field CCD Imager  \citep{NBI2020gbar.conf..127S} and a set of narrow-band filters. Their  parameters (central wavelength and FWHM of the bandwidth) are given in Table~\ref{obs:log} together with other information on the observations. 

Data reduction included all standard steps: bias subtraction, cosmic-ray hits removal, flat-field correction, correction for non-linearity of the detector, background air-glow emission. Individual exposures in each filter were combined with sigma-clipping after correction for the small variations of the differences in flux and PSF FWHM (by fitting Moffat profile to the foreground stars). 
We applied $2\times2$ binning to increase the signal-to-noise ratio and achieve the optimal sampling of the PSF.

One of the two spectrophotometric standards (BD+75D325 and AGK+81D266) was observed during each night in each of the filters immediately before or after the science exposure. We reduced these images in the same way and used them for flux calibration of the images in emission lines. However, the photometric condition had changed between the science and standard observations on February 21, 2018, leading to a non-reliable flux calibration for the \SII images, and no other exposures in this filter were taken on other dates. Therefore, to calibrate \SII\ images, we used SCORPIO-2 \revone{LS} spectra. We also find a good agreement between the fluxes measured from the \revone{LS} spectra and flux-calibrated images in other emission lines. The final continuum-subtracted images were obtained by matching the flux and PSF of the foreground stars in images taken with the filters centered on emission line and adjacent continuum, and subtracting one from another. 

Note that the \Ha{} filter also captures the \NII{}$\lambda6548,6583$ lines. Based on spectral data, we estimate the contribution of these lines to the \Ha{} flux in the studied \HII{} regions to be < 7$\%$.

We produced synthetic maps \SIIHa{} and \OIIIHb{} based on the described photometric data. As a proxy of \Hb{} flux for \OIIIHb{} map (hereafter marked as \Hb{}*) we used \Ha{} flux divided by \revone{$3.1$  (median observed \Ha/\Hb; see Tables~\ref{table_fluxes1},\ref{table_fluxes2},\ref{table_fluxes_custom}). The interstellar extinction towards NGC~2366 is small ($\rm E(B-V) \sim 0.086$~mag, as measured from our spectra), and thus its variations do not significantly affect the calibration}. 
 % $\rm A_V = 0.1$ according to \citealt{Schlafly2011}
%which is confirmed with nearly constant \Ha/\Hb{} value measured with long-slit spectroscopy data, %\textcolor{red}{(2.39 for region 65A, for the rest 2.80 - 3.35)}, 
\revone{Described} maps are shown in Fig.~\ref{fig:lineratiomaps}, where pixels with $\rm{S/N} < 3$ are masked. %We point emission line ratios from the maps integrated within the region's aperture on the BPT diagram (see Fig.~\ref{fig:BPT}).

% The resulting images were calibrated based on the spectral data described in section \ref{subssec:longslit}.

% \begin{figure}
% \includegraphics[width=\linewidth]{Figs/map.pdf}
% \caption{Positions of slits for long-slit spectroscopy of NGC~2366 on BTA 6-m telescope (blue), superimposed on \Ha{} map from CMO 2,5-m telescope. The limits of the star formation regions under study are shown in blue; positions of the cone walls are shown in cyan and correspond to the positions on the Fig.~\Vanya{???}. Pink squares mark the positions of SNRs from the works of \cite{Vucetic2019}, \citet{Ercan2021} and  \cite{Chomiuk2009}; positions of X-ray sources from the work of  \cite{Kaaret2024} are shown by cyan stars. LBV star described in the work of \cite{Drissen1997} is marked by the yellow square. There are also 3 WR stars detected nearby the location of LBV star in the Knot B of Mrk~71 region (not shown on the plot, see \cite{Drissen2000}).}
% \label{fig:map}
% \end{figure}

\section{Analysis}
\label{sec:analysis}

In this paper, our objective is to derive physical properties and chemical abundances of the \HII\ regions across the NGC~2366 and to identify the regions with a particularly high impact of massive mechanical feedback on the surrounding ISM. In this Section, we describe the particular steps to obtain the results.  

\subsection{HII regions\label{subsec:HIIregions}}

We selected 74 \HII{} regions using the astrodendro\footnote{\url{https://github.com/dendrograms/astrodendro}} package applied to the \Ha{} narrow-band image. We set the minimal size of a region equal to $25~\rm{px}$, masking all values below the $5\sigma$ level and merging the regions where flux variations lie within $5\sigma$. Then, we removed all the regions with negative integral flux in \SII{} or \OIII{} images. % as possible false-detected ghosts. 
For \HII{} regions detection we considered area shown in Fig.~\ref{fig:lineratiomaps}a. The selected regions are overlaid on the \Ha{}, \SIIHa{} and \OIIIHb{} maps in Fig.~\ref{fig:lineratiomaps}a,b and c, respectively. %Due to negligible dust attenuation, which is confirmed with nearly constant \Ha/\Hb{} value measured with long-slit spectroscopy data \textcolor{red}{(2.39 for region 65A, for the rest 2.80 - 3.35)}, we use \Ha{} flux divided by $2.79$ (which corresponds to Te=15kK) as an \Hb{} estimate (marked as \Hb{}*). We also limit line ratio maps with $\rm{S/N} \ge 3$ threshold. We point emission line ratios from the maps integrated within the region's aperture on the BPT diagram (see Fig.~\ref{fig:BPT}).

Our catalog of \HII{} regions is served here as a basis for further analysis of the variations of ionization state and metallically across the galaxy. We use the selected \HII{} regions only as references defining the position and borders of the corresponding physically bounded areas. Therefore, we did not aim to derive a complete catalog nor precise borders of the regions (which is impossible with our angular resolution). Qualitatively, our localization of the \HII{} regions is in agreement with what was derived earlier by
\citet{Vucetic2019, Roy1996}.

% \textcolor{red}{all this to results?????? Most regions show relatively low \SIIHa{} ratios, but  regions 8, 24 and 25 show \SIIHa{} $\sim$ 0.4, and large region 29 is remarkably inhomogeneous. 
% Regions 6 and 25 are associated with SNR candidates identified by \cite{Chomiuk2009}. Region 25 shows increased \SIIHa{} with low \OIIIHb{}, which is typical for SNRs. Region 6, on the contrary, shows low \SIIHa{} and high \OIIIHb{}. 
% Several other regions show increased \OIIIHb{} ratio, indicating higher ionization level. Between them, Mrk~71 and its outflow region are clearly outstanding showing high ionization and extremely low \SIIHa{} ratio.  }

\subsection{Small-scale ionized gas kinematics}
 %Within this section we identify high velocity dispersion objects and analyse their kinematics in detail.

%\subsubsection{$I-\sigma$ analysis}
\label{subsec:Isigma}

% $I-\sigma$ diagram was firstly introduced by \cite{Munoz1996}, who compared intensity vs velocity dispersion for each spatial pixel, and later improved by \cite{Moiseev2012}, who changed intensity to integrated flux. The diagram serves as a powerful tool for identification of high energetic power sources. While analysis of dispersion map only is complicated because of regions with low signal-to-noise ratio and diffuse ionised gas component, regions with different physical nature can be clearly separated on the diagram, as it was shown in a number of papers \citep[see][as an example]{Egorov2023}. Following the same approach as in \cite{Egorov2023}, we show $I-\sigma$ diagram for NGC~2366 in Fig.~\ref{fig:isigma}b and corresponding classification map in Fig.~\ref{fig:isigma}a. In Fig.~\ref{fig:isigma}c,d, we provide \Ha{} flux and velocity dispersion maps obtained with FPI observations for comparison.

\begin{figure*}
    \includegraphics[width = \linewidth]{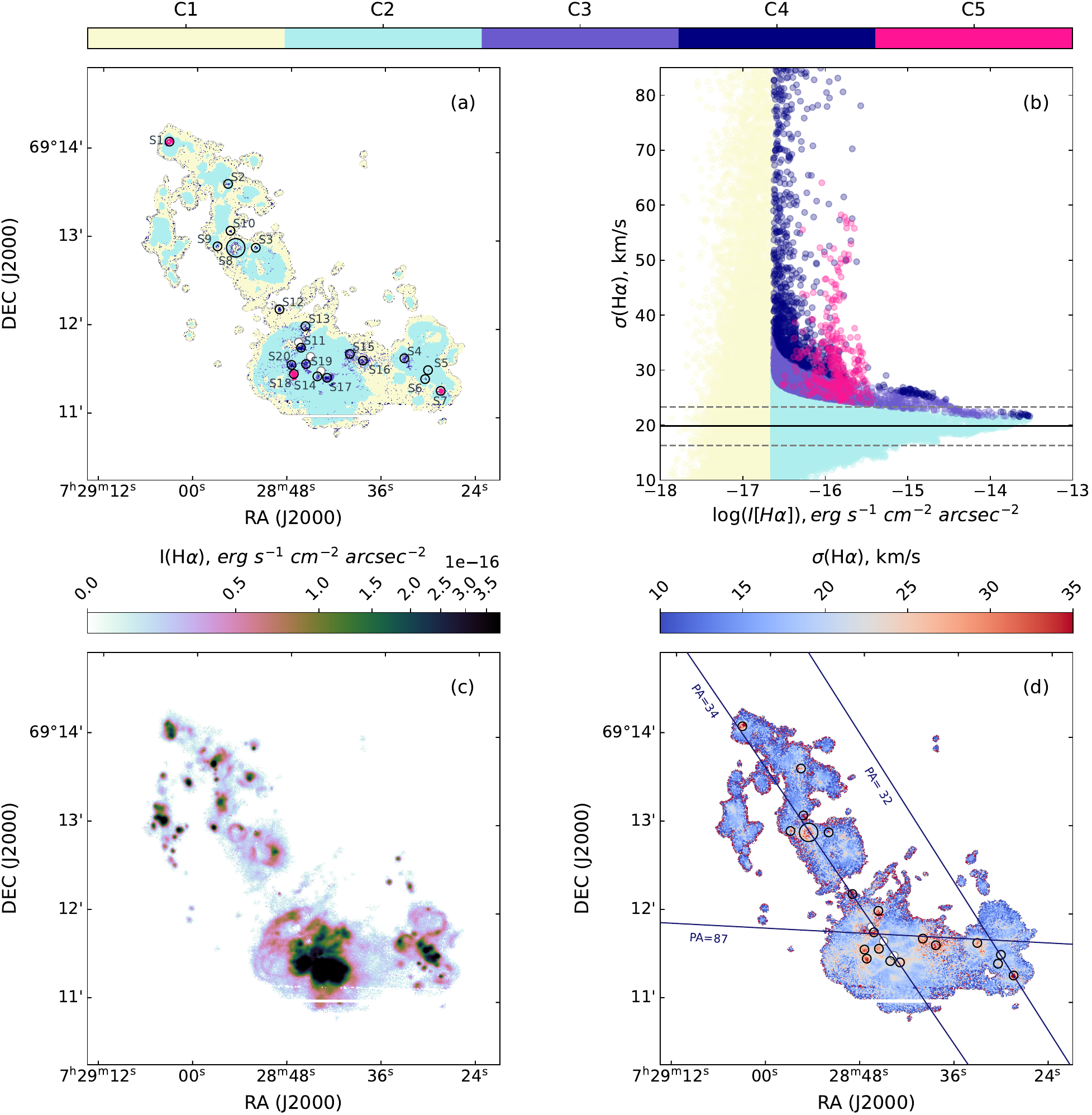}
    \caption{Classification map (a) and $I-\sigma$ diagram (b) coloured according to classification described in the text (see Section~\ref{subsec:Isigma}). C1 refers to DIG, C2 -- \HII{} regions, C3 -- intermediate class with elevated velocity dispersion, C4 -- regions of perturbed ionised gas (such as expanding bubbles and turbulent ISM), C5 --  candidates to individual high mechanical energy sources. \Ha{} image (c) and velocity dispersion map (d) are given for comparison with classified sources. In panel (b) are shown flux-weighted mean velocity dispersion (solid line) and its $1\sigma$ deviation (dashed lines), and in panel (d) slit positions are superimposed. The identified objects (see Section~\ref{subsec:Isigma}) spatially unresolved in our data are indicated with circles, except for the object clearly exhibiting a shell-like morphology, which is marked with a large circle. \revone{In panels (a) and (d) grey circles indicate the masked areas significantly affected by artefacts.}}
\label{fig:isigma}
\end{figure*}

\begin{figure*}

\includegraphics[width=\linewidth]{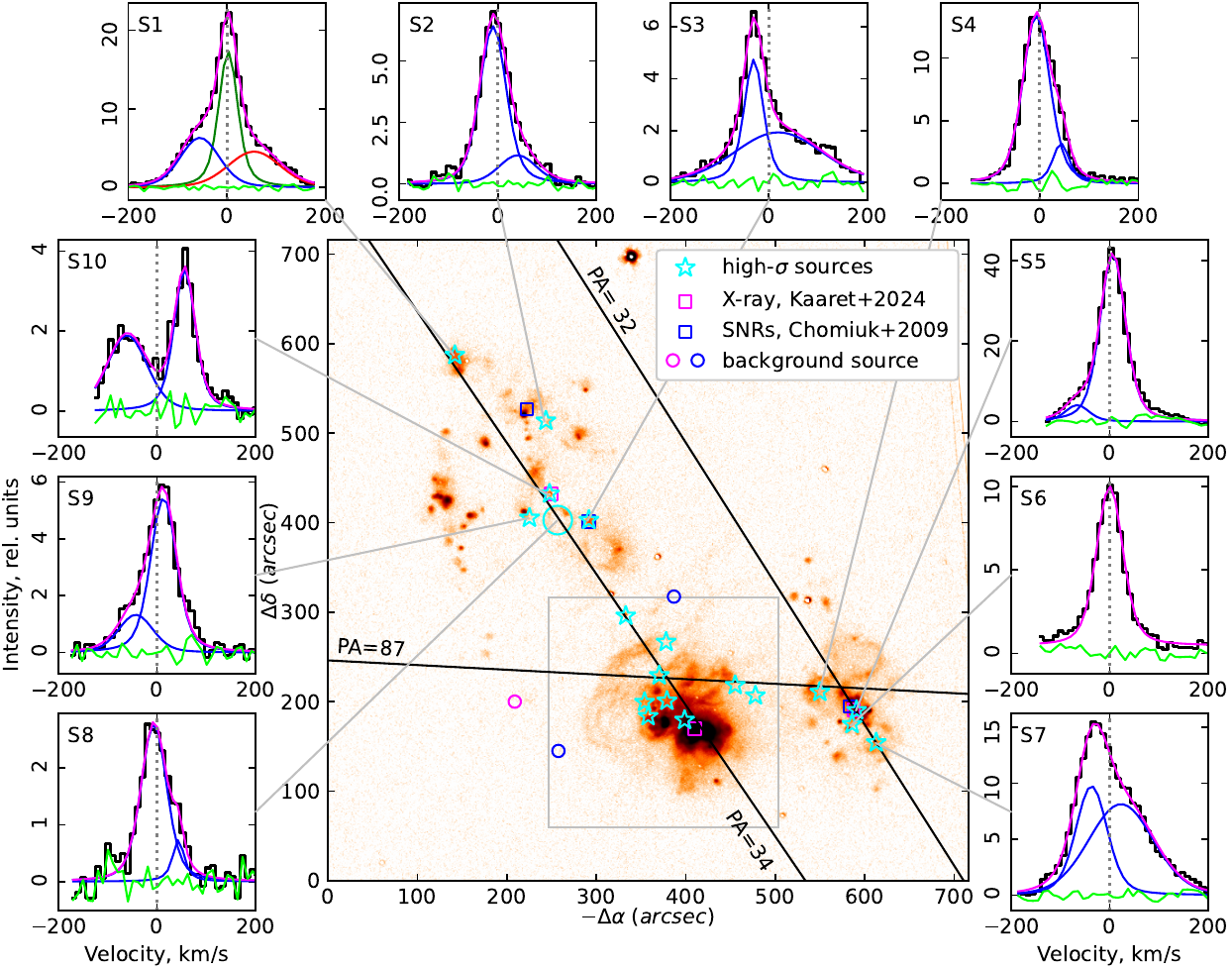}
    \caption{Profiles of regions exhibiting large velocity dispersion identified from FPI data. Central panel shows the intensity distribution in the \Ha{} line with superimposed slit positions (black lines). Locations of regions with turbulent motions revealed in the present work are overlaid with cyan stars and a circle, while the X-ray objects and SNRs from previous works are marked by pink and dark blue squares correspondingly. Side panels show the \Ha{} line profiles in the highlighted regions. The panels also demonstrate the best fit by 1 to 3 Voigt components (pink) and individual components (dark blue) as well as residuals (light green). The central square marks the area shown in Fig.~\ref{fig:profiles_mrk71}}\label{fig:profiles_ngc2366}
\end{figure*}

\begin{figure*}
\includegraphics[width=\linewidth]{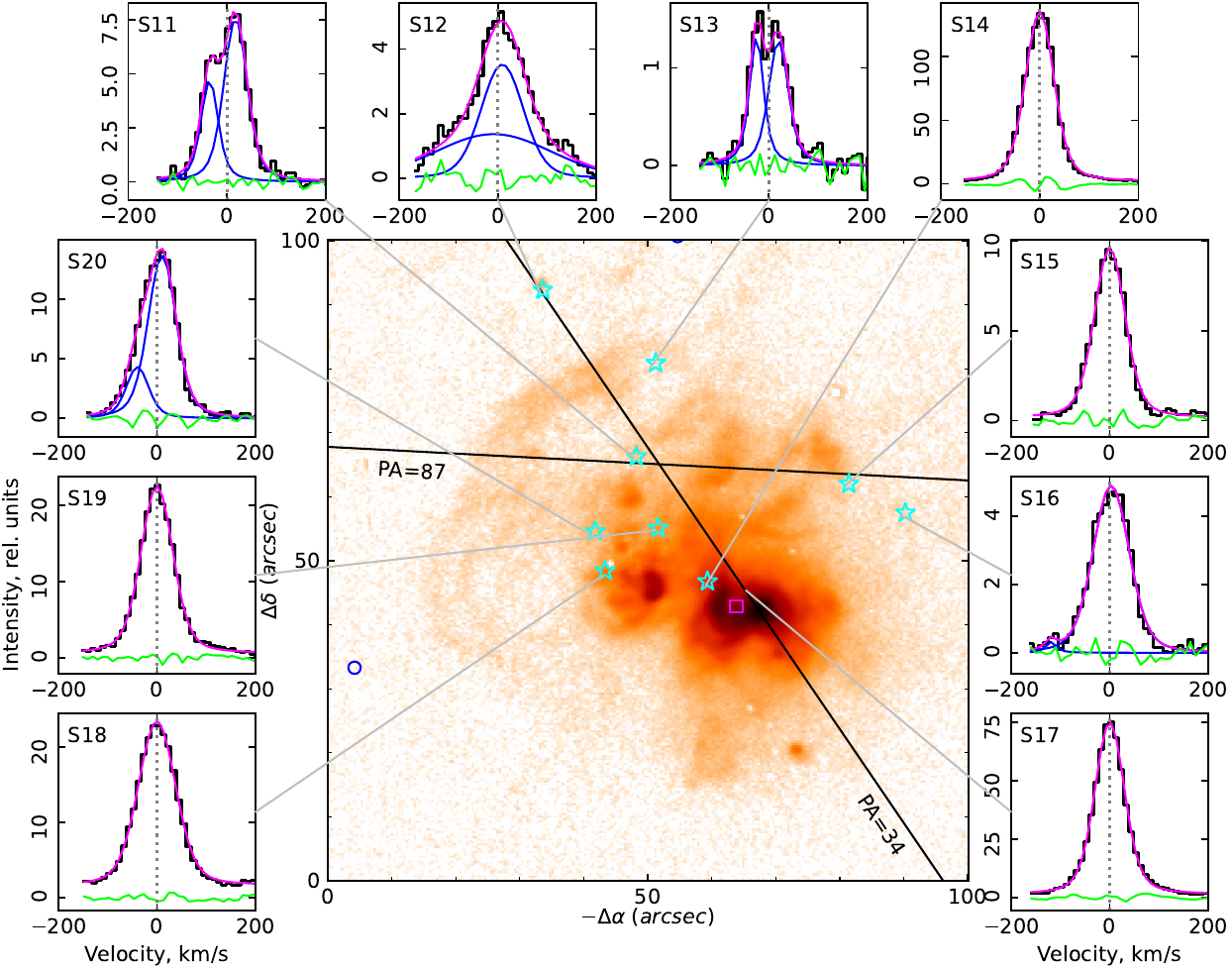}
    \caption{Same as in Fig.\ref{fig:profiles_ngc2366}, but for The Mrk~71 region and its surroundings.}
\label{fig:profiles_mrk71}
\end{figure*}

\begin{table*}
\caption{Parameters of the \Ha{} line profiles shown in Figs.~\ref{fig:profiles_ngc2366},\ref{fig:profiles_mrk71}\label{tab:profiles}}
\begin{tabular}{clcccccccc}
\hline
\# & \HII{} region & RA(J2000) & DEC(J2000) & \multicolumn{2}{c}{First comp.}&\multicolumn{2}{c}{Second comp.} &\multicolumn{2}{c}{Third comp.}\\

& & & & V$_1$,~km~s$^{-1}$ & $\sigma_1$,~km~s$^{-1}$ & V$_2$,~km~s$^{-1}$ & $\sigma_2$,~km~s$^{-1}$ & V$_3$,~km~s$^{-1}$ & $\sigma_3$,~km~s$^{-1}$ \\
\hline
S1 & 1 \revone{(NGC 2366)} & 7:29:03.473 & +69:14:05.05 & -13 & 34 & 44 & 14 & 96 & 48 \\
S2 & 8 & 7:28:47.252 & +69:11:28.29 & -6 & 22 & 44 & 33 &  &  \\
S3 & 25 & 7:28:28.519 & +69:11:17.90 & -5 & 14 & 45 & 84 &  &  \\
S4 & 55 \revone{(NGC 2363)} & 7:28:52.294 & +69:12:53.69 & -48 & 22 & 2 & 11 &  &  \\
S5 & 61 \revone{(NGC 2363)} & 7:28:54.853 & +69:12:53.53 & -116 & 19 & -44 & 21 &  &  \\
S6 & 67 & 7:28:49.178 & +69:12:12.06 & -37 & 21 &  &  &  &  \\
S7 & 69 \revone{(NGC 2363)} & 7:28:46.378 & +69:11:46.18 & -29 & 27 & 31 & 61 &  &  \\
S8 & 24 & 7:28:47.558 & +69:11:34.44 & -14 & 21 & 36 & 5 &  &  \\
S9 & 19 & 7:28:45.715 & +69:11:35.08 & -48 & 30 & 8 & 21 &  &  \\
S10 &  & 7:28:43.015 & +69:11:25.94 & -114 & 36 & 3 & 16 &  &  \\
S11 & 44 & 7:28:38.467 & +69:11:37.87 & -73 & 12 & -21 & 18 &  &  \\
S12 & 33 & 7:28:33.184 & +69:11:39.74 & -20 & 113 & -3 & 38 &  &  \\
S13 &  & 7:28:55.924 & +69:13:36.84 & -63 & 8 & -19 & 15 &  &  \\
S14 & 65 \revone{(Mrk 71)} & 7:28:30.155 & +69:11:31.88 & -27 & 24 &  &  &  &  \\
S15 &  & 7:28:45.842 & +69:12:00.83 & -23 & 26 &  &  &  &  \\
S16 &  & 7:28:40.135 & +69:11:42.30 & -15 & 32 &  &  &  &  \\
S17 & 65 \revone{(Mrk 71)} & 7:28:57.179 & +69:12:54.34 & -19 & 25 &  &  &  &  \\
S18 & 60 & 7:28:55.550 & +69:13:05.03 & -30 & 33 &  &  &  &  \\
S19 & 56 & 7:28:44.251 & +69:11:26.82 & -27 & 27 &  &  &  &  \\
S20 & 56 & 7:28:30.516 & +69:11:25.80 & -77 & 18 & -27 & 23 &  &  \\
\hline
\end{tabular}
\end{table*}

Turbulent gas motions can be driven by different sources injecting mechanical energy to ISM,
such as individual supernovae, WR or LBV stars, or young star clusters. We use here a so-called `$I-\sigma$' diagram to select these sources, assuming that they should reveal high velocity dispersion and exhibit bright \Ha{} emission. The diagram was proposed to efficiently distinguish between kinematically distinct objects, such as expanding bubbles, unresolved supernova remnants and stellar sources with strong winds, from each other and from unperturbed \HII{} regions and DIG using the information on the brightness and  velocity dispersion of ionized gas \citep{Munoz1996, Moiseev2012}. Here, we use its modification and classification proposed by \cite{Egorov2023}. We show $I-\sigma$ diagram for NGC~2366 and the corresponding classification map,  \Ha{} flux and velocity dispersion maps derived from FPI observations in Fig.~\ref{fig:isigma}. 

%On this diagram, we show the pixels with signal-to-noise ratio $\mathrm{S/N>3}$. We measure median flux of all the pixels left \textcolor{black}{($\mathrm{I[H\alpha]_{median}} = 2.35\times 10^{-17}$~erg~s$^{-1}$~cm$^{-2}$~arcsec$^{-2}$)}. 
The median flux of all pixels with signal-to-noise ratio $\mathrm{S/N>3}$ shown   in diagram is $\mathrm{I[H\alpha]_{median}} = 2.35\times 10^{-17}$~erg~s$^{-1}$~cm$^{-2}$~arcsec$^{-2}$.
Those with low \Ha{} surface-brightness (i.e. $\mathrm{I[H\alpha] < I[H\alpha]_{median}}$), we consider as probably dominated by diffuse ionised gas (DIG) or having significant errors of velocity dispersion measurements. These regions are marked as C1 in Fig.~\ref{fig:isigma} (see the colour scale at the top of the Figure).

For regions with $\mathrm{I[H\alpha] > I[H\alpha]_{median}}$, we measured \Ha{} flux-weighted mean intrinsic velocity dispersion and $1\sigma$ deviation of its value: $\sigma_\mathrm{m} \pm \sigma_\mathrm{std} = 19.8 \pm 3.5 \kms$ (solid black and dashed grey lines in Fig.~\ref{fig:isigma}b). This value represents the majority of the regions, which we consider as unperturbed \HII{} regions. All pixels having $\sigma(H\alpha) < \sigma_{norm}$ are classified as C2, where $\sigma_{norm}$ is Gaussian-shape with standard deviation $0.8\sigma_{std}$ centered at $\sigma_\mathrm{m}$ \citep[see details in][]{Egorov2023}. %We note that we define them regions with $\sigma(H\alpha) < \sigma_{norm}$, where $\sigma_{norm}$ are the Gaussian-shape with peak at $\sigma_\mathrm{m}$ and standard deviation of $0.8\sigma_{std}$ \citep[see details in][]{Egorov2023} to take into account the possible uncertainties for low S/N regime.
%The next two classes, C3 and C4, are characterised by the same Gaussian-like shape, but with standard deviation $1.5\sigma_{std}$. 
C3 corresponds to the pixels within $1.5\sigma_{std}$ for corresponding $\rm \log(I)$, while C4 pixels have $\sigma$ higher than this limit. Spaxels marked as C4 correspond to potential expanding bubbles or ionised gas with the dominant impact of turbulent motions, while spaxels marked as C3 typically surround such regions. After visual inspection of the selected sources from C3-C4 class, we separated those with the pixels forming triangle-like shape or diagonal sequence on I-$\sigma$ diagram to C5 class, which is the common feature for the powerful sources of mechanical energy, such as SNRs or WR and LBV stars (see e.g. \citealt{Moiseev2012, Yarovova2023}). %The \Ha{} line profiles of the regions marked as C3-C5 on $I-\sigma$ diagram are shown in Figs.~\ref{fig:profiles_ngc2366},~\ref{fig:profiles_mrk71}. 
Class C5 contains 3 objects -- S1, S7 and S18.  Note that object S8 is the only resolved bubble-like structure with bright rims and elevated velocity dispersion (class C3) in its centre.

% Among the sources distinguished by kinematics, \textcolor{red}{number} coincide with the areas identified in the previous stage during the \Ha{} map analysis.

%\subsubsection{Properties of the regions with high velocity dispersion}

\label{subsec:prop_turb_reg}

We perform multi-component analysis of the selected sources to examine their kinematical properties in detail. The angular resolution of the FPI data is 2\arcsec, which is equivalent to 32 pc. Therefore, in this case we consider `small-scale' kinematics in NGC~2366. In Figs.~\ref{fig:profiles_ngc2366} and \ref{fig:profiles_mrk71}, we present localization of identified sources on the \Ha{} map with their line profiles and multi-component Voigt fitting. Zero velocity corresponds to the first statistical moment for each line profile. Their main kinematic parameters are listed in Table~\ref{tab:profiles}.

% Regions S1-S5, S7-S13 and S20 reveal complex multi-component structure, which indicate probable unresolved SNR or feedback of massive young stars. In contrast, S6, S14, S15 and S17-S19 have only broadened \Ha{} line profiles indicative of probable slow expansion, temperature broadening or presence of shocks. Among the other regions, S3 and S12 showing broad underlying component, which is a sign of presence of young unresolved expanding SNRs or shocks. S7, S10,S13 and S11 reveal clear two-component \Ha line profile indicating expansion or complex movements in these regions. And the most remarkable one is S1 with a complex 3-component line profile.  %The other regions (S2, S4, S5, S8, S9, S20) do not reveal clear evidence of expansion, but turbulent motions there are still significant. 

\subsection{Ionization state of the emission nebulae}
\label{sec:analisys_spec}

\begin{figure*}
\centerline{
\includegraphics[width=\textwidth]{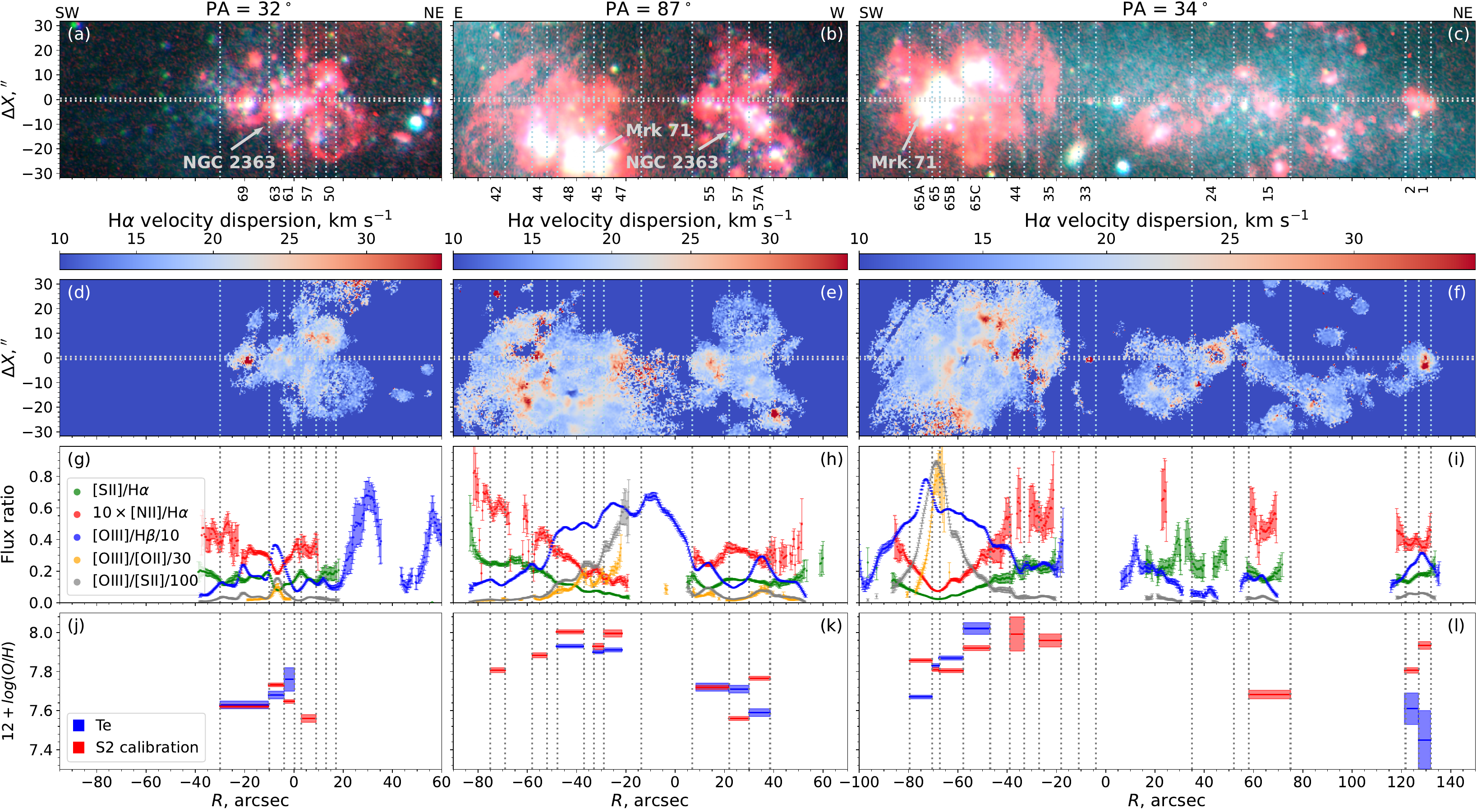}
}
\caption{\revone{Slit locations are superimposed on the false-colour image (B + V + \Ha{}) of NGC~2366 (a,b,c) and on the velocity dispersion map (d,e,f). Distribution of emission line intensity ratios along the slits for \Ha{}, \SII{}, \OIII{} and \NII{} lines can be found on panels g,h,i for position angles PA = 32 (left) and PA = 87$\degr$ (center) and PA = 34 (right). The bottom panels (j,k,i) show the 12+log(O/H) estimations obtained by $T_e$ method and S-calibration.}}
%On this plot, the data is given for slits PA=32 and PA=87. 
\label{fig:slitdistr_pa32pa87}
\end{figure*}

%During the next step, we compare the positions of the \HII{} regions highlighted on the map \Ha{} with the positions of the slits and integrate the parts of the spectra matching the particular \HII{} regions. 
To determine gas ionisation state and metallicity we obtained integrated spectra of \HII{} regions within the borders, \revone{which are presented in Fig.~\ref{fig:slitdistr_pa32pa87} }as vertical grey lines. We note that for regions 1, 44, 45 and 63, slits crosses only the outskirts of the \HII{} nebulae, that can lead to bias in the measured line ratios, and hence temperature and metallicity estimates. %All the other regions are crossed in the middle.
The integral fluxes in the emission lines were measured by fitting a single-component Gaussian to emission line profiles. \revone{In the cases of close or overlapping spectral lines we used multi-component fitting.} The $1\sigma$ uncertainties of the derived fluxes are estimated by repeating measurements for 100 Monte Carlo realizations. The measured fluxes are corrected for reddening using the \textsc{pyneb} {\sc python} package \citep{Luridiana2015} assuming \citet{Fitzpatrick1999} reddening curve. The $E({B-V})$ value is estimated from the Balmer decrement \Ha/\Hb{} consistent with the electron temperature measured for the particular \HII{} region. 
\revone{We did not subtract the underlying stellar absorption lines from  LS spectra because of the negligible contribution of absorbtion. \cite{Chen2023} estimated that the best-fit value for absorption lines equivalent widths (EW) is 0.6 $\pm$ 0.7 \AA~ in the region of Mrk 71, which is indeed much less, than EWs of Balmer emission lines in this region (we estimate $\mathrm{EW_{Ha}}\approx 2000$\,\AA\, for Mrk 71).}
We report the measured fluxes and the corresponding errors for each region in Tables ~\ref{table_fluxes1}, \ref{table_fluxes2}, \ref{table_fluxes_custom}.

Spectra PA=32$\degr$, 34$\degr$, 87$\degr$, obtained in the present work, cover 24 star-forming regions. Slit PA=87$\degr$ covers the region of a conical outflow in Mrk~71 (see Fig.~\ref{fig:morphology}) and crosses the bright region NGC~2363. Slit PA=32$\degr$ also covers NGC~2363 region, crossing its bright center. Finally, slit PA=34$\degr$ crosses the S1 region revealing the complex 3-component \Ha{} line profile and the brightest star-forming region in NGC~2366 -- Mrk~71. % A. 
The structure of the Mrk~71 is complicated, therefore we divided it into 4 regions -- 65, 65A, 65B and 65C. Region 65 corresponds to the knot A in the center of Mrk~71 (it is marked as the grey circle in Fig.~\ref{fig:morphology} \revone{; the region coincides with those analysed in \citealt{Chen2023}}). Region 65B represents the part of the conical outflow close to its tip, and region 65C is located further than the outflow area. Region 65A is located on the opposite side from region 65B with respect to region 65.

% Версия от чатика жпт: Region 65B is located at the apex of the ionization cone, marking the starting point of the gas outflow. Region 65A lies on the opposite side of Region 65 with respect to Region 65B; it includes areas where the broad underlying \OIII{} component remains visible. Region 65C is situated behind Mrk~71, and in this region, we integrate the area where the broad underlying \OIII{} component is still detectable, even outside the cone.

In this section, we present the analysis of gas ionization conditions with both \revone{LS} spectroscopy of the regions and narrow-band maps.

% \subsubsection{Emission line distributions}
Strong emission line ratios are commonly used indicators of gas ionization conditions, helping to disentangle between photoionization by massive stars, hard ionizing radiations, shocks, and mixed mechanisms of gas excitation \citep[e.g., review by][]{Kewley2019}. %Therefore, they are commonly used to trace shocks, ionising photon leakage or hardness of ionising radiation. 
For example, elevated \OIIIHb{} and {\OIII/\OII{}} line ratios are indicative of hard ionizing radiation and possible leakage of ionizing photons from \HII{} regions, while high \SII/\Ha{} is a commonly used indicator of shocks.

In Fig.~\ref{fig:slitdistr_pa32pa87} we show the distribution of the following parameters along the slit: velocity dispersion (d,e,f), line ratios between \Ha{}, \SII{}, \OIII{} and \NII{} for different position angles (g,h,i), oxygen abundance $\mathrm{12+log(O/H)}$ estimated by $T_e$ method and \revone{S-calibration~\citep{Pilyugin2016}} for \HII{} regions crossed by slits (g,k,j; see Sec.~\ref{subsec:gasphase_metall}). To construct these distributions, we %convolved spectra with a rectangular kernel of 5 pixels size ($\approx 2\arcsec$, similar to seeing conditions during the observations) to improve the S/N ratio, and then we 
fitted each emission line pixel-by-pixel along slits with the single-component Gaussian profile. %The underlying continuum was subtracted as a median of pixels, not contaminated by strong lines continuum around considered lines in the same window. %Among the objects identified from the ionized gas kinematics (Sec.~\ref{subsec:Isigma}), the long-slit data cover S1, S4, S5, S7, S8, S11, S12, S15 and S17.

\begin{figure*}
\includegraphics[scale=0.7]{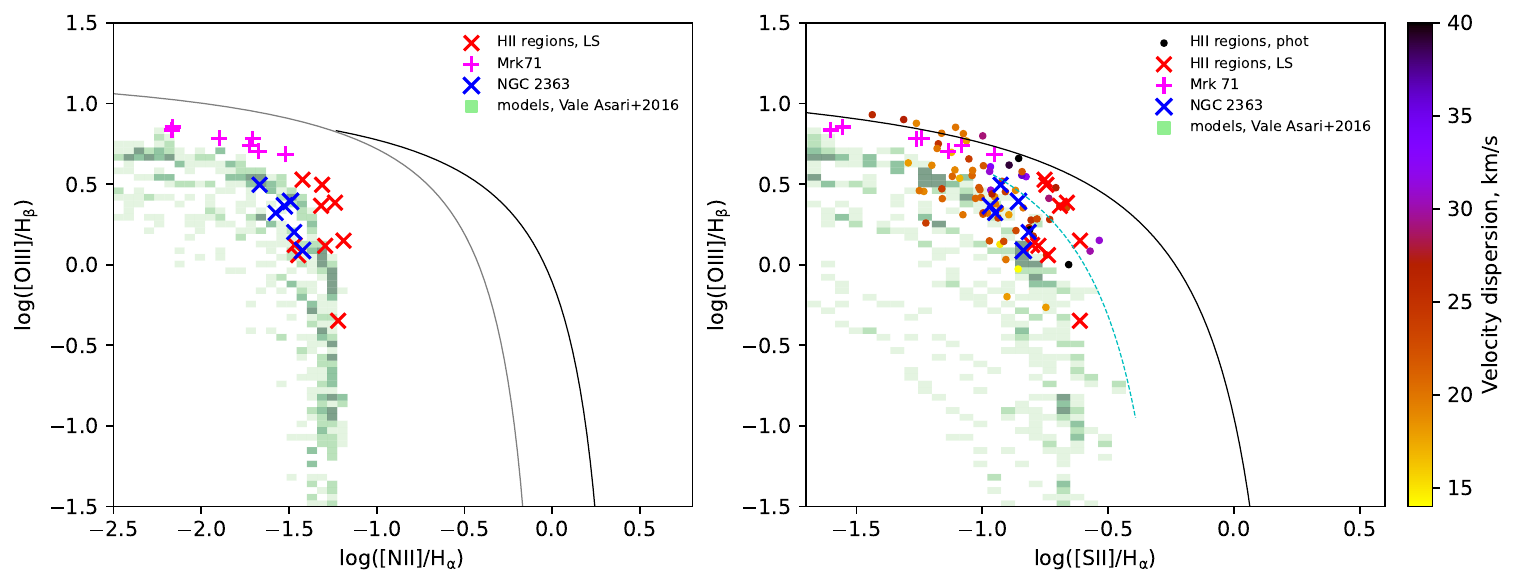}
\caption{Diagnostic BPT diagram, based on relations from \citet{Kewley2001} (black line) and \citet{Kauffmann2003} (grey line), showing boundary between photoionised regions (to the left of the lines) and regions with high contributions of shock excitation. \revone{The models from \citet{Vale2016} for oxygen abundances $\mathrm{12 + \log(\mathrm{O/H}) = 8.0}$ and $\mathrm{N/O} = -1.5$ are shown in the green histogram with the colour-coded density of the photoionisation models grid at a specific line ratios range.} With a cyan dashed line we separate regions where shocks may contribute to ionization (see Section~\ref{sec:analisys_spec}). We also show the \SIIHa{} vs \OIIIHb{} ratio obtained using photometrical data from CMO telescope in filters \SII{}, \Ha{} and \OIII{} together with velocity dispersion for these regions from SCORPIO-2 FPI data (colour scale). \revone{Calibration of the photometrical data is described in Sec.~\ref{subsec:narrowband}}. The positions of studied regions from LS spectroscopy are overlaid: Mrk~71, %+ conical outflow of ionised gas
(pink '+'), NGC~2363 (blue 'x') and other regions (red 'x'). 
}
\label{fig:BPT}
\end{figure*}

To further analyse ionization state of the studied regions, we construct diagnostic BPT diagrams \citep{BPT, Veilleux1987} and show them in Fig.~\ref{fig:BPT}. To separate the photoionization regions from the regions, where shock waves contribute significantly, we use the demarcation curves \NIIHa{} vs \OIIIHb{} and \SIIHa{} vs \OIIIHb{} (black lines) from the works of \citet{Kauffmann2003} and \citet{Kewley2001}. Although the given models are classical, they do not describe well low-metallicity conditions such as in NGC~2366 (see Sec.~\ref{subsec:est_gas_par}). To make comparison more robust for the metal-poor ISM, we add to the plot the photo-ionisation models from \citet{Vale2016} as a green histogram, constructed for the metallicity of $\rm 12+\log(O/H)=8$ and $\rm \log(N/O) = -1.5$, that is consistent with our measurements (see Table.~\ref{tab:Te_Oabund}). For reference, we also plot the ratio \SIIHa{} vs \OIIIHb{} from our photometrical data for the distinguished \HII{} regions (see Sec.~\ref{subsec:HIIregions}) as circles with colour-coded velocity dispersion to highlight perturbed regions (the datasets and the calibration is described in Sec.~\ref{subsec:narrowband}). We showed the values for both photometric and \revone{LS} datasets. We also codded by different symbols the regions Mrk~71 and NGC~2363 in Fig.~\ref{fig:BPT}). %, which we identify as cWR later in this work. 
%The diagrams are shown in Fig.~\ref{fig:BPT}. 
We empirically defined the demarcation line between photoionised regions and those potentially affected by other ionization mechanisms at metallicity of NGC~2366 as envelope of the \cite{Vale2016} models  on \SIIHa{} vs \OIIIHb{} diagram. This line is shown in Fig.~\ref{fig:BPT} by cyan dashed line. The regions considered in our work generally reside within the selected photoionisation model grid, although for those located beyond that cyan line we cannot exclude contribution of shocks to the cumulative excitation. 

% We do not see any regions in the diagram that clearly indicate ionization by a mechanism other than photoionization. However, the region corresponding to cWR lies further than the other regions for which spectral data were analyzed, which may indicate some contribution of shock waves to the ionization of the nebula. 

\subsection{Gas-phase metallicity}
\label{subsec:gasphase_metall}

According to presented BPT diagrams, most of the regions crossed by long slits show predominantly photoionization mechanism of excitation. Therefore, we can use `direct' $T_e$-based method and strong emission line empirical calibrations to estimate the oxygen abundance in these regions.
% As our previous analysis shows photoionisation as the dominant excitation mechanism, we can perform chemical analysis for 21 of 24 regions (spectra of regions 24, 33 and 50 does not show reliable S/N in the emission lines for further diagnostics). 

For 15 regions, we measured auroral line \OIII$\lambda$4363, necessary for accurate estimates of the electron temperature. %Therefore, we derive these parameters where possible.
We measured $T_e$(OIII) corresponding to zones with doubly-ionized oxygen, $\rm O^{2+}$, from \OIII{}$\lambda$4363/$\lambda$5007+4959 line ratios using  \textsc{pyneb} {\sc python} package (see \citealt{Luridiana2015}). To estimate the electron temperature in zones dominated by single ionized oxygen, we used the empirical relation $T_e(\mathrm{O}^+)=0.7T_e(\mathrm{O}^{2+})+3000$K from \cite{Garnett1992}. 

For the $T_e$ measurements, we integrated the spectra within the area where \OIII$\lambda4363$ is detectable at the level $>3\sigma$ in each pixel along the slit for each region. We assumed that such $T_e$ is representative for high-ionization ($O^{2+}$) zone of the entire region. Similarly, we limited the integration area by the region with measurable \OII$\lambda 3727$ for the subsequent metallicity estimates. 
%For $T_e$ determination, we integrated area  where the \OIII$\lambda4363$ line is visible. To estimate the oxygen content, accordingly, parts of the spectra that do not contain the line \OII$\lambda 3727$ were not taken into account during the summation procedure. 
The integrated spectra, including zoom-in on \OIII$\lambda$4363, are shown in Figs.~\ref{fig:first10} and \ref{fig:second10}. 

We estimated $n_e$ in \textsc{pyneb} package based on \SII$\lambda6731$/\SII$\lambda 6731$ ratio. All studied regions, except 65, 65A and 65B, corresponding to Mrk~71, have $n_e$ close to the low density limit for such method ($\sim30~\rm{cm}^{-3}$), in which precise estimation is not possible due to the model degeneration. For further calculations of abundances, we fixed the density for those regions at the value of 30~$\rm cm^{-3}$. Note that metallicity calibrations are almost insensitive to changes of $n_e$ in the low-density regime typical for \HII{} regions. For 65, 65A and 65B, we obtained correspondingly $n_e = 250, 100, 150\, cm^{-3}$.

Using the determined temperature and electron density, we calculated the oxygen and nitrogen abundances in \textsc{pyneb}: $\mathrm{O^{2+}}$ metallicity was derived from the \OIII$\mathrm{\lambda}5007$/\Hb{} flux ratio and $\mathrm{O^{+}}$ metallicity from the \OII$\mathrm{\lambda}3727$/\Hb{} ratio. The O/H metallicity was then estimated as $\mathrm{(O^{2+}}+\mathrm{O^{+}})/\mathrm{H^{+}}$. $\mathrm{N/O}$ was estimated as $\mathrm{N^+/O^+}$ based on \NII$\mathrm{\lambda}6583$/\Hb{} ratio. The results are shown in Table~\ref{tab:Te_Oabund} and \revone{discussed in Section~\ref{subsec:est_gas_par}}.

For 21 regions, we measured integrated fluxes in the \OIII{}, \Hb{}, \NII{} and \SII{} lines necessary for the empirical \revone{S-calibration~\citep[see eqs.~6,7 in][]{Pilyugin2016}}. Although this method is generally consistent with the $T_e$-method (as it is calibrated against \HII{} regions with measured $T_e$), it is less precise and can be more sensitive to variations in ionization conditions producing deviations from the `typical' \HII{} regions. \revone{In particular, \SIIHa{} ratio is sensitive to shocks and DIG, and thus their presence can significantly affect the measured metallicities.}   %This method is less reliable, as it does not retrieve physical condition of ISM directly, but it can be used in \HII{} regions where the positions on the BPT diagram are similar to the ones by which it was calibrated. We assume that the same position on the diagram reflects similar physical conditions.

For both $T_e$ and S methods, uncertainties were estimated using Monte-Carlo method with 100 iterations, taking into account variations in line fluxes and temperature within the errors.% and density. 

% We were able to estimate $T_e$, $\mathrm{12+log(O/H)}$ and $\mathrm{log(N/O)}$ for three spectral regions of PA=32$\degr$, six regions of PA=87$\degr$ and six regions of PA=34$\degr$. Our estimations for the studied regions are listed in the table \ref{tab:Te_Oabund}. For most of the regions $\mathrm{12+log(O/H)}$ vary from 7.6 to 7.9. Obtained oxygen abundance for \HII{} regions in the NGC~2366 disc is significantly lower than that measured by \cite{Roy1996} ($\sim$ 8.2). They measured the metallicity based on the narrow-band images with the O3N2 empirical calibrator, which is insensitive in case of low metallicities \citep{Marino2013}, that explains the inconsistency between our results.

% Based on our measurements, we notice a significant difference in the oxygen abundance between the Mrk~71 region and the associated ionized gas outflow region, and the rest of the galaxy (NGC~2363 and regions 1,2). In Sec.~\ref{sec:discussion} we discuss the reasons of possible metallicity variations in the galaxy

% We also note, that region 1 lies apart from the rest, showing higher temperature and lower oxygen levels. Since this region lies on the upper boundary on the BPT diagram (see Fig.\ref{Figs/BPT_2366_met=8_singlepoints.pdf}), we can assume the contribution of impact ionization. This is also consistent with our assumption that this region is ionized by a Wolf-Rayet star, which we discuss in detail in section \textcolor{red}{in discussion} In the case of shock ionisation, using the $T_e$ method becomes incorrect. 

 \begin{table*}
 \centering
 \caption{$\mathrm{T_e}$ and $\mathrm{12+log(O/H)}$ determined by direct method.}
\begin{tabular}{l|l|l|l|l|l}
\hline
Slit                  &region & $N_{reg}$    & $\mathrm{T_e, 10^3~kK}$ & $\mathrm{12+log(O/H)}$ & log(N/O) \\ \hline
\multirow{6}{*}{PA=87$\degr$} &\multirow{3}{*}{outflow} &  48 & 14.11 $\pm$ 0.17 & 7.93 $\pm$ 0.02 & -1.40 $\pm$ 0.02 \\
                       && 45 & 14.23 $\pm$ 0.26 & 7.90 $\pm$ 0.02 & -1.43 $\pm$ 0.04 \\
                       && 47 & 14.6 $\pm$ 0.2 & 7.91 $\pm$ 0.01 & -1.35 $\pm$ 0.03 \\  \cline{2-2}
                       &\multirow{6}{*}{NGC~2363}& 55 & 14.3 $\pm$ 0.5 & 7.72 $\pm$ 0.02 & -1.43 $\pm$ 0.03 \\
                       && 57 & 13.8 $\pm$ 0.6 & 7.71 $\pm$ 0.02 & -1.45 $\pm$ 0.02 \\
                       && 57A & 16.0 $\pm$ 0.5 & 7.59 $\pm$ 0.02 & -1.31 $\pm$ 0.03 \\\cline{1-1}
                          % \hline
\multirow{3}{*}{PA=32$\degr$}  && 61 & 14.6 $\pm$ 1.3 & 7.76 $\pm$ 0.06 & -1.62 $\pm$ 0.06 \\
                        && 63 & 15.6 $\pm$ 0.3 & 7.68 $\pm$ 0.02 & -1.50 $\pm$ 0.04 \\
                        && 69 & 15.9 $\pm$ 0.8 & 7.63 $\pm$ 0.03 & -1.51 $\pm$ 0.03 \\\cline{1-2}
                          % \hline
\multirow{6}{*}{PA=34$\degr$}   &\multirow{2}{*}{NGC~2366}&1 & 22.3 $\pm$ 2.0 & 7.4 $\pm$ 0.1 & -1.35 $\pm$ 0.24 \\
                        &&2 & 18.1 $\pm$ 2.7 & 7.61 $\pm$ 0.08 & -1.48 $\pm$ 0.16 \\\cline{2-2}
                        &\multirow{4}{*}{Mrk~71}& 65 & 16.12 $\pm$ 0.05 & 7.83 $\pm$ 0.01 & -1.394 $\pm$ 0.012 \\
                        && 65A & 18.31 $\pm$ 0.09 & 7.67 $\pm$ 0.01 & -1.516 $\pm$ 0.014 \\
                        && 65B & 14.92 $\pm$ 0.06 & 7.90 $\pm$ 0.01 & -1.446 $\pm$ 0.025 \\
                        && 65C & 13.5 $\pm$ 0.4 & 8.02 $\pm$ 0.03 & -1.66 $\pm$ 0.05 \\

                       \hline
\end{tabular}
\label{tab:Te_Oabund}
\end{table*}

\begin{figure*}
\includegraphics[width=0.95\linewidth]{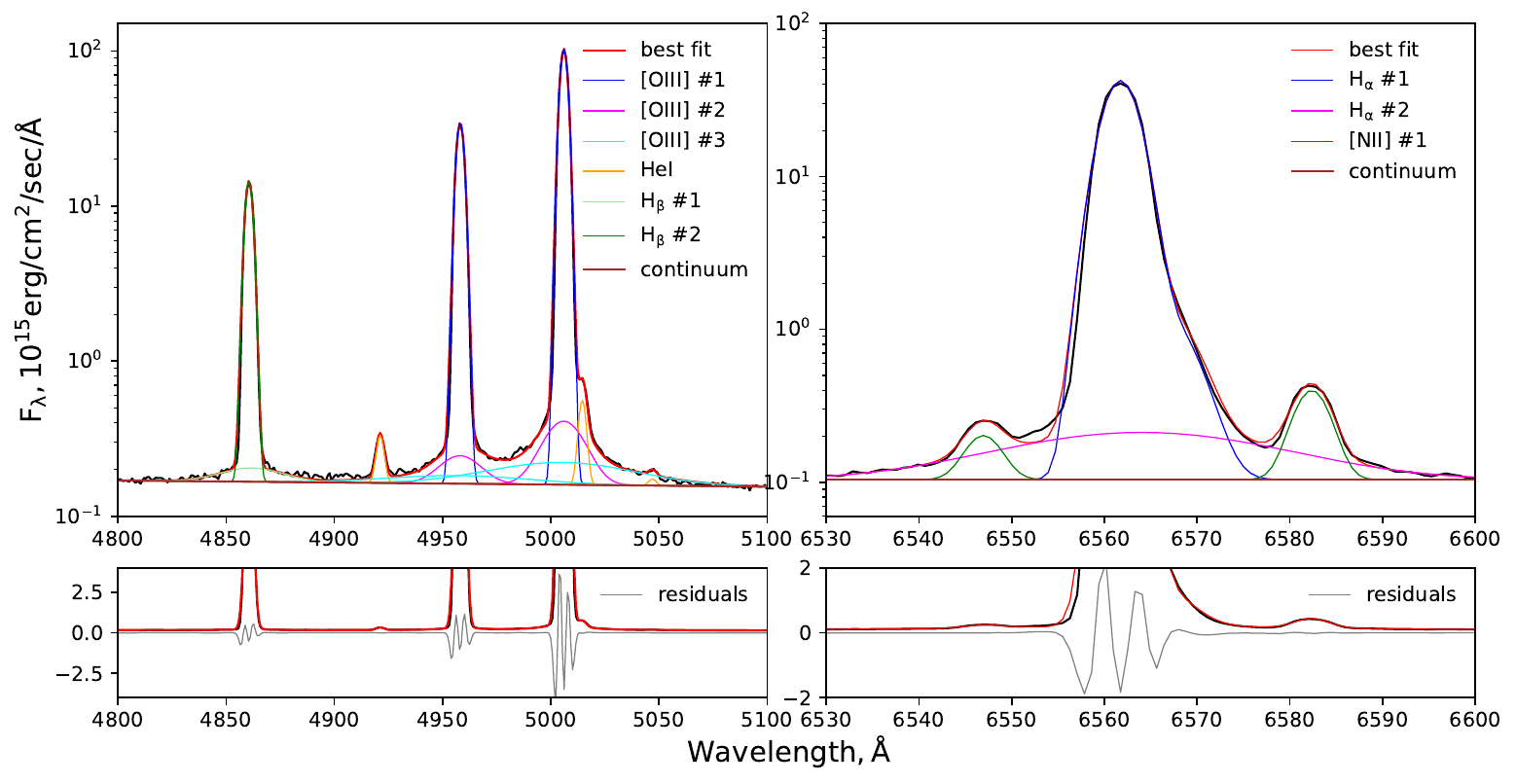}
\caption{We highlight two regions of spectrum 65 (central part of Mrk~71) around the \Ha{} and \OIII{} spectral lines, demonstrating broad components. \revone{To obtain the reliable model, all the lines in each spectral window were fitted simultaniousely.} The top two panels show the spectra in log scale (black), together with best-fit models (red) and components of these models. The bottom panels show the same regions of the spectra in linear scale together with propagation errors (gray). 
}
\label{fig:66}
\end{figure*}

\revone{LS} spectroscopy data from the Mrk~71 region is the most complex, therefore we consider it separately. The most remarkable details in the spectra are the broad components in \Ha{} and \OIII{} lines, which cannot be fitted by single Gaussians, like the rest of the lines. We show the parts of the spectrum 65 around the \OIII{} and \Ha{} lines, as well as best-fit models and separate fitted components for each line on Fig.~\ref{fig:66}. Our best-fit models suggest the presence of two broad components in addition to the main one in \OIII{} line. \Ha{} line shows one broad component, while its main component is strongly asymmetric due to non-Gaussian instrumental profile in the red part of the SCORPIO-2 spectrum \citep{KatkovChilingarian2011}. This asymmetry is significant only for the brightest Mrk~71 region as it is visible at $\sim2$\% level. To compensate for this effect, we fit two Gaussians to this component and consider their sum for the flux measurements. %(see Fig.~\ref{fig:66}). 
For Te, oxygen and nitrogen abundance determination, we used only main components of \Ha{} and \OIII{}. Two central spectra show high level of excitation, according to the criterion $\mathrm{O^+/(O^+ + O^{2+})<0.1}$ from \cite{Izotov2006}, therefore we estimate Ionization Correction Factor (ICF) for $\mathrm{12+log(O/H)}$ for these two regions. For both regions, ICF does not exceed 1\% of $\mathrm{(O^+ + O^{2+})/H}$. The measured values of electron temperature and abundances are given in Table~\ref{tab:Te_Oabund}.

\section{Results and discussion}
\label{sec:discussion}

Using \revone{LS} and 3D spectroscopy data, we study number of \HII{} regions of the dwarf metal-poor galaxy NGC~2366 revealing complex kinematics, elevated velocity dispersion or complex line profiles with FPI spectroscopy. \revone{LS} spectra were obtained for the Mrk~71 region, areas coinciding with the ionized gas outlow in Mrk 71, NGC~2363, and 9 objects showing several components in FPI data (see Fig.~\ref{fig:profiles_ngc2366}). Below we discuss the possible nature of these objects as well as our estimates of physical parameters of studied \HII{} regions. 

\subsection{Study of WR/SNR candidates}
\label{sec:wr}

Using scanning FPI with a spectral resolution power of $R\sim16\,000$, we found 20 kinematically distinct regions exhibiting broad or multi-component line profiles in H$\alpha$ line. These regions are named as S1--S20 in Figs.~\ref{fig:isigma} and Table~\ref{tab:profiles} and overlaid as cyan stars in Figs.~\ref{fig:profiles_ngc2366},~\ref{fig:profiles_mrk71}. Here, we discuss on the possible nature of these sources.
%All the regions show perturbed kinematics exhibiting either complex multi-component structure or elevated velocity dispersion in \Ha{} line profile. %Among others, we highlight region S1 with three-component \Ha{} line profile. 

We put all the regions on the BPT diagram to find a possible impact of shocks to their ionization. While all the objects lay inside the photoionisation zone for the solar metallicity models, the studies of other low-metallicity dwarf galaxies show \citep[e.g][]{Egorov2021, Gerasimov2022, Gerasimov2024} that comparison with the photoionization models tailored to the measured metallicity \citep[e.g.][]{Gutkin2016, Vale2016} provide more robust constrains on the excitation mechanisms for particular sources. % in low metallicity environment. 
According to \cite{Vale2016}'s models, the objects S1, S2, S3, S11 lay outside the photoionisation zone (see the cyan demarcation curve on Fig.~\ref{fig:BPT}). Therefore, we conclude that for these 4 objects, photoionization cannot explain their observed line ratios, and other excitation mechanisms (such as shocks) can contribute to their ionization. Additionally, we highlight the objects S3 and S5 as they are SNR candidates according to \cite{Chomiuk2009, Vucetic2019, Ercan2021} (see also Fig.~\ref{fig:profiles_ngc2366}) and S10, which has an X-ray counterpart \citep{Thuan2014, Kaaret2024}. For the rest objects, we do not have any additional information to reveal their nature and therefore it remains unclear.

The bright \HII{} region S1 is of particular interest due to its 3-component \Ha{} profile (see S1 in Fig.~\ref{fig:profiles_ngc2366}). We suggest that the bright component in the \Ha{} line may be associated with the central source, while the two broad spectral components with the expanding bubble around it. We estimate the expansion velocity of the bubble as $55~\kms$ and upper limit of age $< 1$~Myr (the shell is unresolved in our images, so we can estimate only upper limit of its size). These measurements are indicative of the presence of a high-energy central source, such as a WR star or SNR. The object lies outside the photoionization zone on the BPT diagram (see Fig.~\ref{fig:BPT}), therefore we can assume the contribution of shocks. The \revone{LS} spectrum of the S1 region shows a faint \HeII$\lambda$4686 emission line with $\mathrm{EW}=14.5$\AA{} and $\mathrm{S/N}=3$. 
% The associated uncertainty was estimated as
% $$\sigma_l = \sigma_c\sqrt{N+EW/\Delta},$$
% where $\sigma_c$ is the standard deviation in the nearby continuum, N is the number of pixels of FWHM of the \HeII{} line, EW is the equivalent width of the line and $\Delta$ is the dispersion of the spectrograph in \AA~px$^{-1}$. 
\HeII{}~$\mathrm{\lambda4686}$ line is a common tracer of WR stars. It also indicates a high temperature of the region, which is consistent with our $T_e$ estimation for this region, giving the highest value throughout the galaxy (22 kK, see Tab.~\ref{tab:Te_Oabund}). All the above makes this object a candidate to a WR star. %The neighbouring \HII{} region 2, also crossed by the slit, does not reveal the \HeII{} line nor a multi-component structure. The ratios of the emission lines \OIIIHb{} and \SIIHa{} in these two spectra also differ, while \NIIHa{} ratios are similar. This discrepancy indicates that \HII{} regions 1 and 2 have different nature. 
Currently, there are several detected WR stars in NGC~2366 (see \citealt{GonzalezDelgado1994, Drissen2000}) and one LBV star \citep{Drissen1997}. However, all of these previously detected objects are located in the Knot B of Mrk~71 region and do not coincide with S1. Thus, we discovered a new WR candidate at the northern part of NGC~2366. \revone{The follow-up observations are required to confirm detection of \HeII$\lambda$4686 emission line with higher S/N.}  

Other notable objects are S2 and S11, which have two-component \Ha{} profiles, and S3, which reveals broad \Ha{} line component. They all lay outside the photoionisation zone on the BPT diagram, according to the models of \cite{Vale2016}. Unlike S1, these objects do not show emission in \HeII{}, which is common for SNRs. Object S3 is mentioned in the previous papers as SNR candidate \citep{Chomiuk2009, Vucetic2019}, and our kinematical study confirms its status. We consider the other two objects -- S2 and S11 -- as a new SNR candidates. Another previously known SNR candidate coincides with our S5 object, which exhibits asymmetric \Ha{} line profile. \revone{However, it resides within the photoionisation zone on the BPT diagram with $\text{\SIIHa{}} = 0.11-0.15$, and thus its interpretation as SNR is unclear. We note, that the value can be significantly affected by nearby \HII{} regions.} %S5 is noted in previous works and has asymmetric \Ha{} line profile, however this object lay inside photoionisation zone on the BPT diagram, which makes it controversial.

One more remarkable object is S10. It shows a two-component \Ha{} line profile with a distance between them equal to $117~\kms$. \revone{It also has an X-ray counterpart with one of two possible sources XMMU J072855.4+691305 or XMMU J072851.7+691255 (\citealt{Thuan2014}). While the former has luminosity $6 \times 10^{36}\rm\ erg\ s^{-1}$, that is comparable with high-mass binary stars in the Galaxy, the latter is faint and marginally blended with the first one. It can possibly originate from a binary system or extended hot plasma. All this makes S10 a candidate to SNR or an energetic X-ray binary star.} 

There are some interesting objects which nature is not clear. One of them is S12, which reveals clear broad-line \Ha{} component. This is very intriguing as broad emission line components are often interpreted a clear kinematical sign of shocks. However, its spectrum does not reveal any metal emission lines necessary to confirm its ionization state on the BPT diagram. S7 is also very interesting object as it reveals very bright broad-line \Ha{} component and shows a clear triangle-like shape on the I-$\sigma$ diagram. It shows that this object is clearly very energetic, but it lays within the photoionisation zone with $\rm T_e = 16\,000~K$. Therefore, this object is unlikely an SNR, and there is not enough information to consider it as a WR star.
\revone{The other object S18, laying on the diagonal sequence of the I-$\sigma$ diagram (see Fig.~\ref{fig:isigma}), belongs to the class C5, which attracts particular attention to their nature. However, it also does not stand out on the BPT diagram. We also note that neither S7 nor S18  do not coincide with any of the known stellar associations revealed by \citet{Kaaret2024}. For now, we do not have enough spectral information and therefore future studies are required to reveal their nature.}

Summing up, based on the FPI and \revone{LS} data analysis, we conclude that region S1 is a candidate to a Wolf-Rayet star surrounded by an expanding bubble, S2 and S11 are new SNR candidates, and S10 is an SNR or an X-ray binary star candidate. For S3, we confirm its status as a strong SNR candidate.

\subsection{Chemical abundances of HII regions}
\label{subsec:est_gas_par}

For 15 \HII{} regions, we estimate $\mathrm{T_e,~ n_e,~ 12+log(O/H)}$ and $\mathrm{log(N/O)}$ using the $T_e$ method, including Mrk~71, NGC~2363 and the brightest Northern regions (see localization in Fig.~\ref{fig:slitdistr_pa32pa87} and values in Table~\ref{tab:Te_Oabund}).
%48, 45 and 47, corresponding to the conical outflow of ionised gas from Mrk~71; 55, $\mathrm{57_{PA=87}}$, 57A, 61, 63 and 69, residing in NGC~2363; 1 and 2, corresponding to the bright \HII{} regions in the Northern part of NGC~2366 disc, demonstrating complex multi-component structure in FPI data; and four regions, corresponding to Mrk~71 (65 - 65C). 
For all these and other 6 regions we additionally calculated  $\mathrm{12+log(O/H)}$ using the empirical strong-line S-calibration.

Mrk~71 has the most exciting spectrum in the galaxy exhibiting a complex structure in the \Ha{} and \OIII{}$\mathrm{\lambda5007}$ lines. We measured a large number of emission lines corresponding to different ionization states of H, He, O, S, N, Ne, Cl, Ar, Fe (see Fig.~\ref{fig:PA34_starform}). For central bright region of Mrk~71 knot A (region 65), we measured the oxygen abundance $\mathrm{12+log(O/H)}$ = 7.83 $\pm$ 0.01. The obtained value is consistent with the previous works, estimating oxygen abundance for Mrk~71 (7.8 in \citealt{Noeske2000}; 7.85 in \citealt{Izotov2011}; 7.89 $\pm$ 0.1 in \citealt{Chen2023}). We also note that the chemical structure of Mrk~71, obtained in our work, is not homogeneous, but shows a gradient from $\mathrm{12+log(O/H)}=7.67$ to 8.02 (see Table~\ref{tab:Te_Oabund}). Recent joint analysis of integrated data in the central region of Mrk~71 with Hershel/PACS and Keck/KCWI data \citep{Chen2023, MendezDeldago2024} demonstrated controversial results for the same set of data about the presence of temperature variations. Our study finds a steep temperature gradient along the slit from $13500 \pm 500~\rm K$ (region 65C) to $T=18310\pm90~\rm K$ (region 65A). Our estimates indirectly support the possible existence of temperature variations inside region 65 as well. However, the quality of the data (seeing $2\arcsec$) does not allow us to resolve the region 65, which is considered in the mentioned works. High spatial resolution data can provide a new perspective to this question. %\textcolor{red}{which?}.

% For central part of Mrk~71, as well as for the region of outflow, $\mathrm{12+log(O/H)}$ obtained by S2 method is consistent with those obtained using $\mathrm{T_e}$ mrthod. For other regions the agreement is not so precise and the discrepancy reaches 0.2 dex, exceeding this value for only one region, corresponding to new-detected WR candidate. This discrepancy may be due to the small S/N ratio in the case of the \OIII$\mathrm{\lambda}4363$ and \OII{}$\mathrm{\lambda}3727$ lines. Also, S2 method can show variations from the Te method, since it does not take into account the temperature variations across the \HII{} regions, therefore being less reliable. Despite the discrepancies, the S2 method also clearly indicates higher metallicity values in the Mrk~71 and outflow regions than in the rest of the galaxy regions. $\mathrm{log(N/O)}$, measured by Te method for 15 regions, does not show any clear gradient.

The oxygen abundance for the Mrk~71 region obtained in this paper is close to the estimates obtained for the regions coinciding with outflow (regions 48, 45, 47). 
%, which confirms the connection of the outflow with the Mrk~71. 
However, for more distant regions of the galaxy, the oxygen abundance values are found to be significantly lower: NGC~2363 shows lower values of oxygen abundance than Mrk~71; %and the outflow in both long-slit spectra and by both $T_e$ and S-calibration methods, 
region 15 also show\revone{s} reduced value obtained with S-calibration; for region 1 our results are controversial: $T_e$ method shows reduced value, while S-calibration does not. $\mathrm{log(N/O)}$, measured with $T_e$ method for 15 regions, does not show any clear gradient. We do not see the change in N/O content between Mrk 71 and the rest of the galaxy. We also note that \revone{region 1, having $12+\log\mathrm{(O/H)} < 7.6$, has a high uncertainty} due to the lowest S/N values and elevated velocity dispersion (see reg. S1 in Fig.~\ref{fig:profiles_ngc2366}), which could indicate the potential contribution of shock waves and make the estimates by empirical S-calibration (and possibly -- by $T_e$ method) unreliable. As we discussed earlier, we consider this region to be a candidate for WR star.

In \cite{Roy1996} it is shown that the gradient in NGC~2366 is flat with the value $8.19 \pm 0.14$. In contrast, \cite{James2016} shows chemical inhomogeneity in Mrk~71 on scales of less than 50 pc. Our results are consistent with the findings of \cite{James2016} for Mrk 71, and we also show that there are metallicity variations throughout the galaxy as well. The data obtained are too sparse to measure the radial metallicity gradient, but we show that the oxygen abundance in NGC~2366 is not uniform and lies in the range of $\mathrm{12+log(O/H)}=7.6-8.0$. \revone{\cite{KewleyDopita2002} showed that the \NII/\OIII{} calibration strongly depends on ionisation parameter, which varies across NGC 2366. \cite{Marino2013} also showed that O3N2 calibration becomes insensitive to metallicities below $\mathrm{12+log(O/H)}\sim8.2$. The same effect is expected for \NII/\OIII{} relation. Finally, the estimates of metallicities from the photometric measurements as was performed in early studies can also be uncertain due to contamination of \NII\ images by the adjucent bright \Ha\ line. These effects could explain the discrepancy between our results and those from the work of \cite{Roy1996}.}

Thus, for the first time we measured the metallicity using $T_e$ method outside the brightest region of NGC~2366. The difference in metallicity can be explained by the chemical enrichment of the ISM around Mrk~71 \revone{as a result of the outflow}. \revone{Since Mrk71’s starburst has been active for several million years and has already produced a kiloparsec-scale outflow, along with confirmed supernovae and Wolf-Rayet stars, the enrichment of the interstellar medium with heavy elements appears to be a natural consequence of stellar feedback processes.} A similar effect was found, for example, in a galaxy J1044+0353, also showing a giant outflow (\citealt{Martin2024, Peng2023}).
\revone{Our findings support the idea that dwarf galaxies should not always be considered as chemically homogeneous (the same conclusion was obtained in the work  \citealt{James2020}); this challenges the assumption of instantaneous and homogeneous mixing often adopted in chemical evolution models.}

% For the listed regions, we obtained temperature estimations in the range from 14 to 22 kK, with the upper $\mathrm{T_e}$ values corresponding to the area of new-found cWR, though having the highest uncertainty. The metallicity estimations throughout the galaxy vary from 7.4 to 8, with the values \~7.8-7.9 obtained for the central region of Mrk~71 and the outflow associated with it. This agreement once again confirms the close connection between the outflow and the Mrk~71 region that ionizes it. The lower oxygen abundance in the remaining regions, demonstrated by both the $T_e$ and S2 methods, indicates an enrichment of the Mrk~71 region and the adjacent outflow with heavy elements, which may be associated with active star formation in this region.

% Before the present work, the oxygen abundance of the Mrk~71 region has been studied many times, showing average values in the range 7.8 - 7.9. Thus, our measurements are consistent with the literature. For the rest of the galaxy, metallicity measurements were carried out only once, and the obtained values diverge greatly from our estimates (\~8.2 vs \~7.6-7.7), which is explained by the difference in methods: O3N2 empirical calibrator, used for previous estimations, is insensitive in the low-metallicity cases, while $\mathrm{T_e}$ method proves to be more reliable.

\subsection{Kinematical properties of Mrk~71 and the ionised gas outflow}

The conical outflow from the Mrk~71 region was identified by \citet{Roy1991} and carefully described by \citet{vanEymeren2009} and \cite{Micheva2019}. It reveals underlying broad components in
 \OIII{} and \Ha{} lines, which were extensively explored in the previous papers, with several physical mechanisms proposed by \cite{Binette2009} and \cite{Komarova2021}. In this paper, we do not aim to study in detail the kinematics of this region and delve into its physical nature (this will be a focus of the forthcoming paper by Vasiliev et al. in prep), but we briefly discuss the primary features seen in our spectral data. 

 Broad outflow component remain over-resolved in our FPI data because it's narrow wavelength range around H$\alpha$ line prevents its disentangling from the continuum. This explains generally low velocity dispersion measured in this region (Fig.~\ref{fig:isigma}). Meanwhile, about half of the regions with locally elevated velocity dispersion are located at the edges of the outflow (see Figs.~\ref{fig:lineratiomaps}c and \ref{fig:isigma}), and some of them can be associated with its impact onto the ISM in the dense clumps. In our \revone{LS} data, we see broad components in both \Ha{} and \OIII{} lines of Mrk~71. Moreover, these components are present in all 4 regions 65-65C, indicating the outflowing component present throughout all these regions. Thus, the extent of the broad component turns out to be at least $\sim 530$~pc. Our best-fit model implies the presence of two broad components in the \OIII{} line profile (see Fig.~\ref{fig:66}) \revone{with the velocity dispersion $\sim 550$~$\kms$ and $\sim 2000$~$\kms$, respectively (see Table~\ref{tab:kin_par}), compared to 1100~$\kms$ measured by \citet{Micheva2019}}. %That also can be explained by non-thermal nature of the component and thus the different non-gaussian emission line distribution \citep[see turbulent mixing layers model introduced by][as an example]{Binette2009}. 
All of these details once again emphasize \revone{the complex kinematic and mostly unclear structure of the Mrk~71} region and require deep IFU spectroscopic observations of the entire region. 

In this work, conical outflow was crossed by long slit observations twice: PA=87$\degr$ intersects the bright region near the base of the cone (knot A); PA=32$\degr$ intersects the ionized gas filaments located at $\sim 2000$ pc from knot A. Being at a significant distance from the bright part of the cone, the gas intersected by the PA=32$\degr$ slit demonstrates an increased ratio of \OIII/\Ha{} emission lines, which may indicate an underestimation of the size of the ionization region (see also \OIII/\Hb{} map in Fig.~\ref{fig:lineratiomaps}). The further discussion on the possible extent of the outflow and its mass-loading factor will be presented in Vasiliev et. al. (in prep.).

\section{Summary}
\label{sec:summary}
We present a comprehensive spectroscopic analysis of the nearby low-metallicity dwarf galaxy NGC~2366 using high-resolution Fabry–Perot interferometry and long-slit spectroscopy obtained with the 6-m BTA telescope, and narrow-band images centred on emission lines from the 2.5-m CMO telescope. Our main findings are as follows:

\begin{itemize}
\item We identified 20 compact regions with elevated \Ha{} velocity dispersion, most of which exhibit multi-component line profiles. This feature is indicative of mechanical feedback from supernova remnants, Wolf–Rayet stars, or young star clusters perturbing the ISM or blowing out expanding bubbles around these objects. Two of these objects coincide with previously known SNR candidates, whereas three are associated with known X-ray sources. 

\item We classify two objects (S2 and S11), demonstrating peculiar kinematics and contribution of non-photoionization mechanism of excitation as new candidates for SNR. We also confirm the status of previously known SNR candidate (S3) and classify S10 as SNR or an X-ray binary star candidate.

\item We report the discovery of a previously unclassified Wolf–Rayet star candidate in the northern part of NGC~2366. This region shows a distinct triple-component \Ha{} profile (indicative of the presence of an expanding bubble) and faint \HeII{}$\lambda4686$\AA\, emission, along with the highest electron temperature (Te $\sim 22000$K) measured in the galaxy.

\item For 15 \HII{} regions, we measured gas-phase chemical abundances using the direct $T_{\rm e}$ method: $\mathrm{12+log(O/H)} = 7.6$--$8.0$ and $\mathrm{log(N/O)} \sim -1.3$ to $-1.6$. Estimates made using the empirical S-calibration for another 8 regions are consistent with these measurements. 

%\item In the starburst region Mrk~71, we find evidence for a chemical gradient and spatially extended outflow, consistent with broad \OIII{} and \Ha{} line components across multiple zones. The metallicity of the outflowing gas is in agreement with the central region, supporting a common origin.

\item We found an elevated value of $\mathrm{12+log(O/H)}$ in Mrk~71 and the related outflow compared to the other \HII{} regions in the galaxy. We interpret this as the result of potential enrichment of the ISM by metals caused by the outflow in Mrk~71. Overall, Mrk~71 region exhibit measurable spatial variations of $T_e$ and oxygen abundance in the region. 
\end{itemize}

\section*{Acknowledgements}
% SAO
The authors thank Roman Uklein,  and Dmitry Oparin  who performed the BTA observations. 
\revone{We thank the anonymous referee for their careful and detailed analysis of the work and for the comments that we found very useful.}
Observations with the SAO RAS telescopes are supported by the Ministry of Science and Higher Education of the Russian Federation. The renovation of telescope equipment is provided within the national project ``Science and Universities''.
%Zasov (need to check)
The study was conducted under the state assignment of Lomonosov Moscow State University.
% % Oleg
OE acknowledges funding from the Deutsche Forschungsgemeinschaft (DFG, German Research Foundation) -- project-ID 541068876. 
% Astrodendro
This research made use of astrodendro, a Python package to compute dendrograms of Astronomical data (http://www.dendrograms.org/). 
% Astropy
This research made use of Astropy (\url{http://www.astropy.org}) a community-developed core Python package for Astronomy \citep{astropy:2013, astropy:2018}. 
% HST
Partly based on observations made with the NASA/ESA Hubble Space Telescope (PID 13041, PI: B. James), and obtained from the Hubble Legacy Archive, which is a collaboration between the Space Telescope Science Institute (STScI/NASA), the Space Telescope European Coordinating Facility (ST-ECF/ESA) and the Canadian Astronomy Data Centre (CADC/NRC/CSA).
% goszadaniye
%The study was conducted under the state assignment of Lomonosov Moscow State University.

\section*{Data Availability}
The data underlying this article will be shared on reasonable request to the corresponding authors.

\bibliographystyle{mnras}
\bibliography{article} % if your bibtex file is called example.bib

\appendix
\section{Long-slit spectra of the \HII{} regions}
\label{sec:appendixA}

\begin{figure*}
\includegraphics[width=\linewidth]{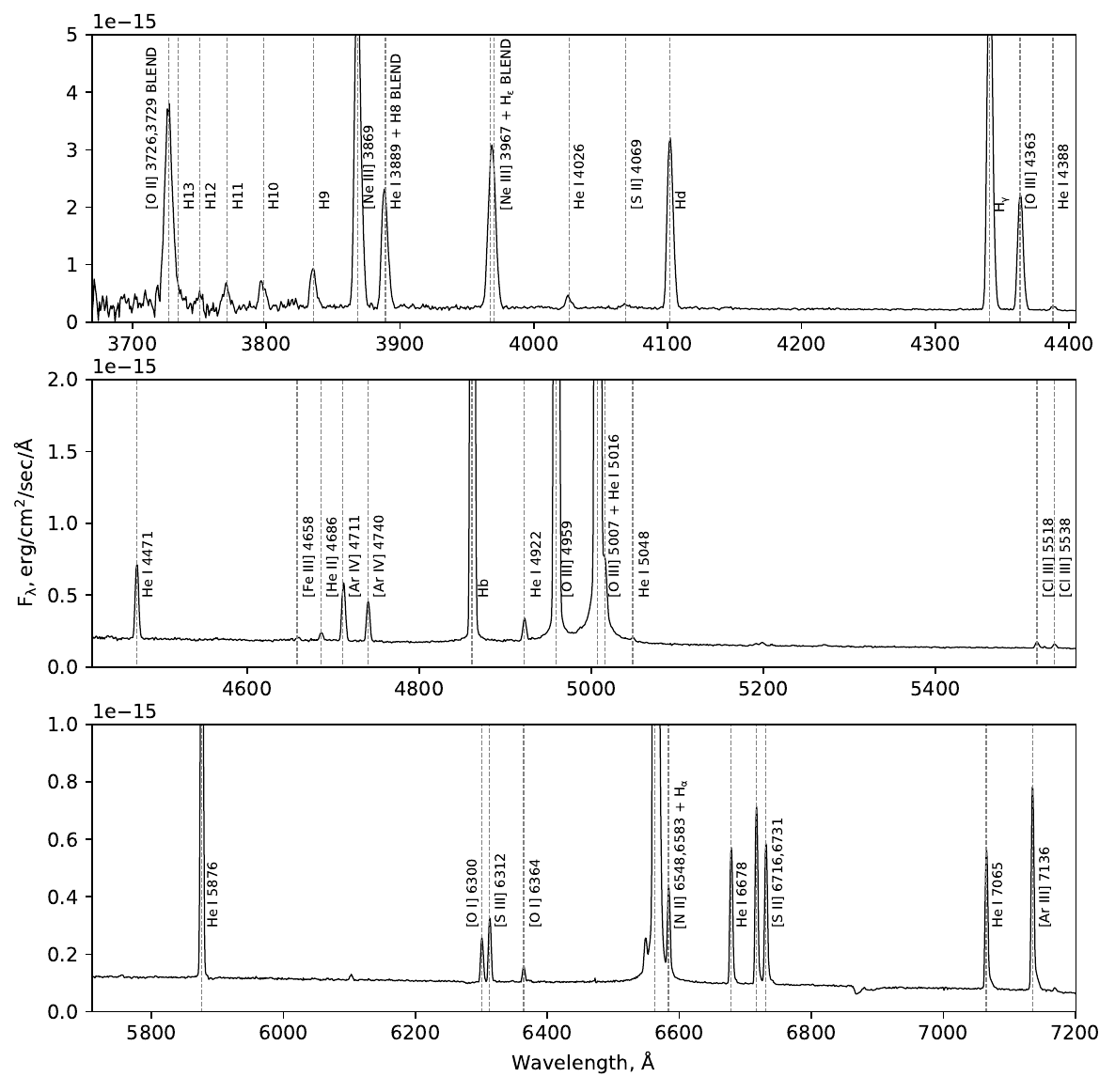}
\caption{Long-slit spectrum of star-forming region Mrk~71 crossed by the slit PA=34$\degr$. Spectrum shows emission lines of Balmer series and \OIII, \OII, \OI, \NII, \SII, \SIII, \HeI, \FeII, \ClIII, \ArIII, \ArIV{} emission lines. The multi-component structure is clearly visible in the lines \OIII$\mathrm{\lambda4959,5007}$ and \Ha. 
}
\label{fig:PA34_starform}
\end{figure*}

\begin{figure*}
\includegraphics[width=0.95\linewidth]{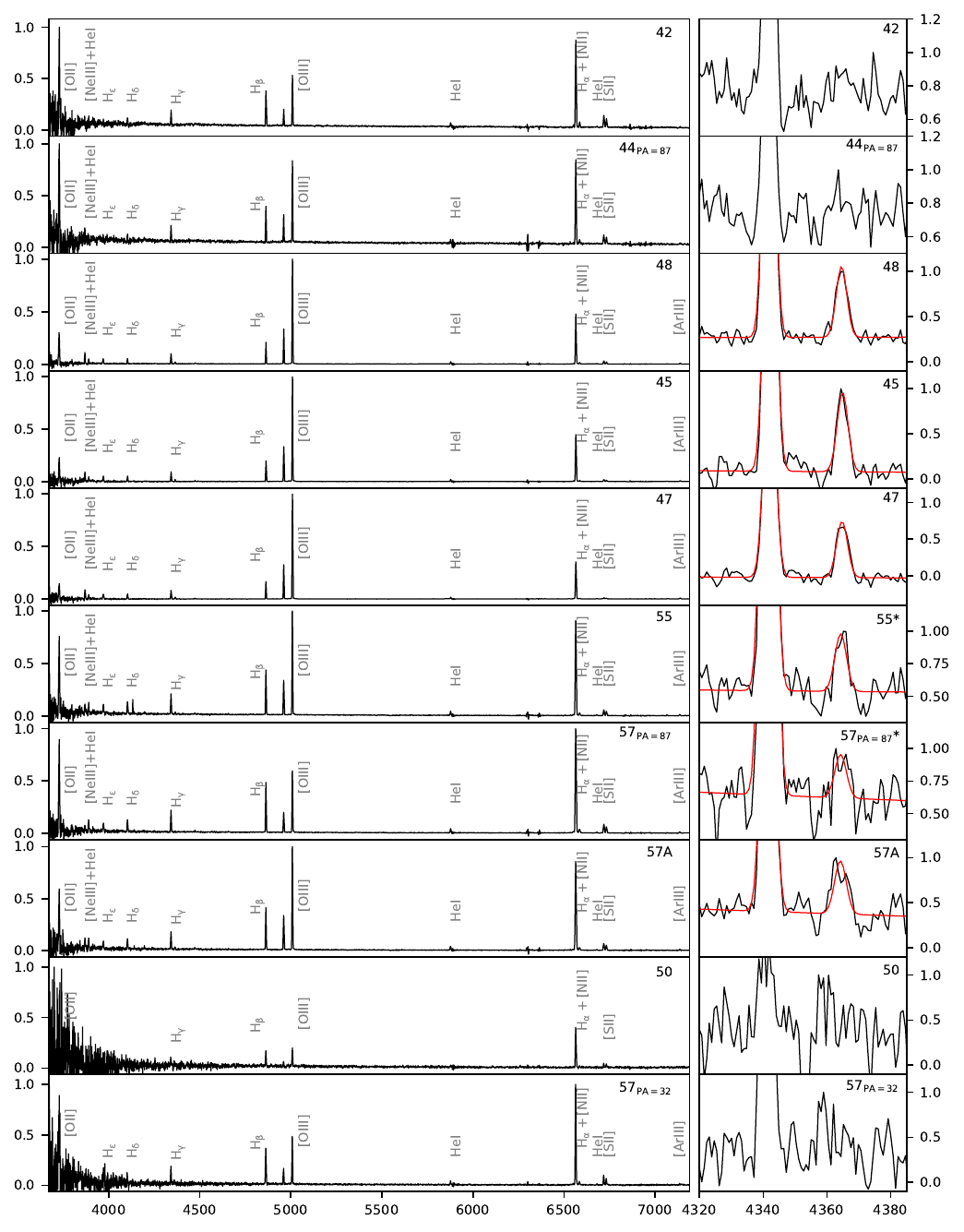}
\caption{Long-slit spectra of \HII{} regions crossed by the slits PA=87$\degr$ and PA=32$\degr$, corresponding to the summation limits for $\mathrm{12+log(O/H)}$ estimation. Each spectrum is supplemented by a panel with an enlarged region of the \OIII{} $\lambda4363$ line with fit shown for those spectra that exhibit $S/N_{[OIII]} > 3$. To determine the temperature of the $T_e$ regions using the $T_e$ method, some of the spectra were summed within narrower limits corresponding to the visibility range of the $\lambda4363$ line. Such areas are marked with the symbol '*'. The region number for each spectrum (and separately for the panel with the \OIII{} line) is signed in the upper right corner.
}\label{fig:first10}
\end{figure*}

\begin{figure*}
\includegraphics[width=0.95\linewidth]{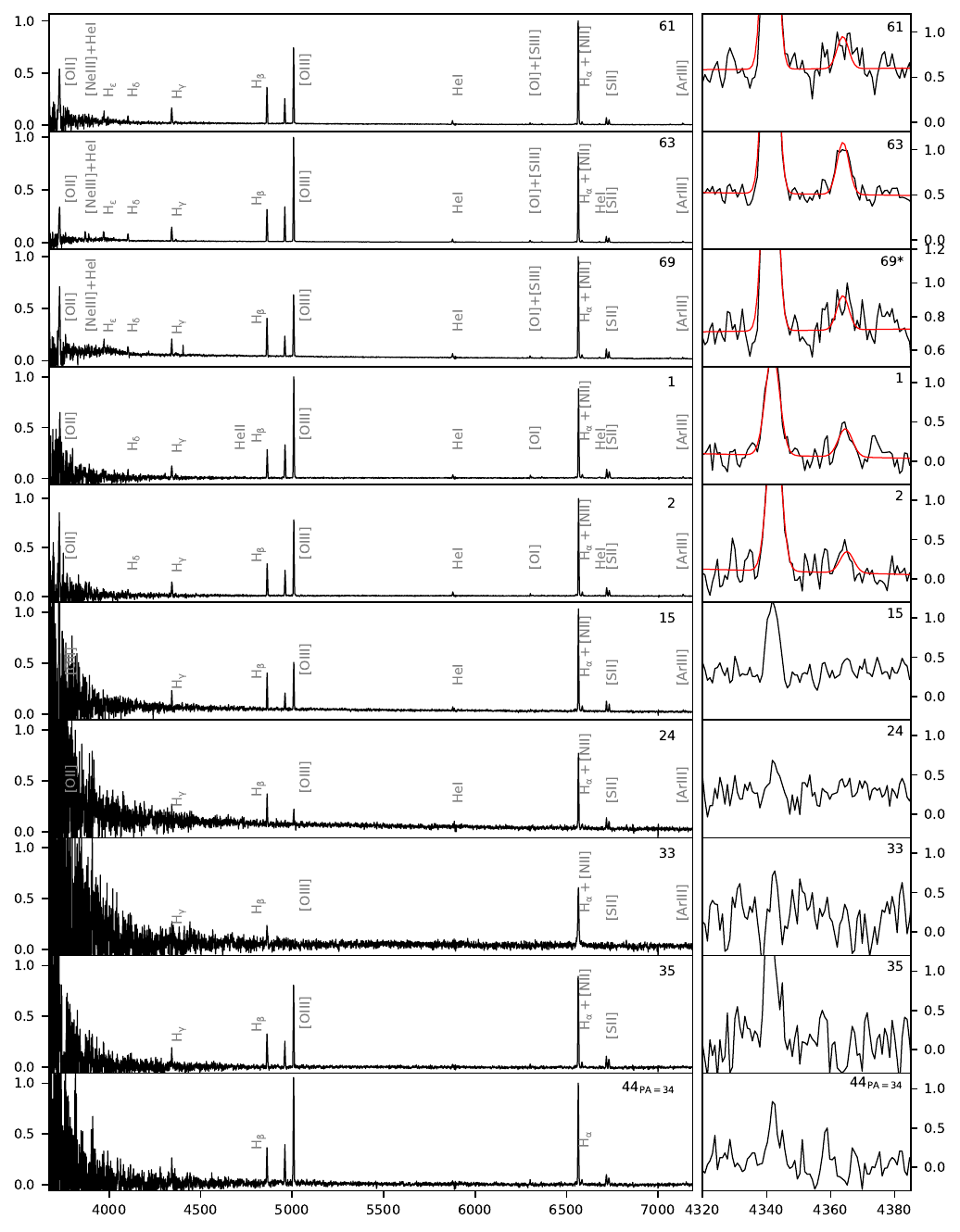}
\caption{Same as in Fig.\ref{fig:first10} for \revone{LS} spectra of \HII{} regions crossed by the slits PA=32$\degr$ and PA=34$\degr$. 
}
\label{fig:second10}
\end{figure*}

Here we present tables with ionization line fluxes of the spectra described in the paper (Tabs.~\ref{tab:Te_Oabund_1}, \ref{tab:Te_Oabund_2},\ref{tab:emission_lines_66_66C}); \revone{LS} spectrum of the central part of star-forming region Mrk 71 (reg. 65, Fig.~\ref{fig:PA34_starform}); Table with kinematical parameters of \OIII{} and \Ha{} lines for regions 65 -- 65C (Tab.~\ref{tab:kin_par}); and spectra of 20 regions, crossed by our \revone{LS} observations (Figs.~\ref{fig:first10}, \ref{fig:second10}).

\begin{table*}
\centering

\small
\onecolumn
 \caption{\label{table_fluxes1}Table presents observed \Ha/\Hb{} without reddening correction and emission line fluxes, corrected for reddening, obtained from the \revone{LS} spectra of \HII{} regions 48, 45, 47, 55, $\mathrm{57_{PA=87\degr}}$, 57A in NGC~2366. For all lines except \OIII$\lambda$4363, we present the fluxes relative to the \Hb{} line corresponding to the spectra used for oxygen abundance determination. For \OIII$\lambda $4363 line, we present \OIII$\lambda $4363/\OIII$\lambda$5007 ratio corresponding to the spectra used for $T_e$ determination (see section \ref{subsec:gasphase_metall}).\label{tab:Te_Oabund_1}
}
\begin{tabular}{l|l|l|l|l|l|l}
\hline
region & 48 & 45 & 47 & 55 & $\mathrm{57_{PA=87}}$ & 57A \\
\hline
\Ha/\Hb{} (obs. values)  & 2.96 $\pm$ 0.01 & 2.96 $\pm$ 0.02 & 2.84 $\pm$ 0.01 & 3.09 $\pm$ 0.01 & 3.15 $\pm$ 0.02 & 3.17 $\pm$ 0.02 \\
\hline
\OII$\lambda3726,3729$ & 1.78 $\pm$ 0.06 & 1.35 $\pm$ 0.10 & 1.08 $\pm$ 0.05 & 1.89 $\pm$ 0.11 & 2.45 $\pm$ 0.08 & 1.71 $\pm$ 0.09 \\
\NeIII$\lambda3869$ & 0.413 $\pm$ 0.017 & 0.406 $\pm$ 0.021 & 0.482 $\pm$ 0.020 & 0.184 $\pm$ 0.016 & 0.107 $\pm$ 0.009 & 0.235 $\pm$ 0.017 \\
\HeI$\lambda3889$ & 0.187 $\pm$ 0.009 & 0.183 $\pm$ 0.013 & 0.217 $\pm$ 0.010 & 0.184 $\pm$ 0.022 & 0.202 $\pm$ 0.012 & 0.225 $\pm$ 0.016 \\
$\mathrm{H_{\epsilon}}$ & 0.253 $\pm$ 0.010 & 0.304 $\pm$ 0.015 & 0.330 $\pm$ 0.011 & 0.228 $\pm$ 0.010 & 0.179 $\pm$ 0.009 & 0.196 $\pm$ 0.016 \\
$\mathrm{H_{\delta}}$ & 0.256 $\pm$ 0.007 & 0.270 $\pm$ 0.015 & 0.308 $\pm$ 0.008 & 0.288 $\pm$ 0.009 & 0.261 $\pm$ 0.008 & 0.268 $\pm$ 0.011 \\
$\mathrm{H_{\gamma}}$ & 0.465 $\pm$ 0.005 & 0.466 $\pm$ 0.010 & 0.479 $\pm$ 0.007 & 0.469 $\pm$ 0.006 & 0.458 $\pm$ 0.006 & 0.443 $\pm$ 0.008 \\
\OIII$\lambda4363*$ & 0.0821 $\pm$ 0.0020 & 0.088 $\pm$ 0.004 & 0.112 $\pm$ 0.003 & 0.0404 $\pm$ 0.0027 & 0.0199 $\pm$ 0.0024 & 0.056 $\pm$ 0.003 \\
\HeII &-&-&-&-&-&-\\
\OIII$\lambda5007$ & 4.838 $\pm$ 0.018 & 5.06 $\pm$ 0.03 & 6.06 $\pm$ 0.03 & 2.316 $\pm$ 0.012 & 1.226 $\pm$ 0.007 & 2.472 $\pm$ 0.016 \\
\HeI$\lambda5875$ & 0.110 $\pm$ 0.002 & 0.112 $\pm$ 0.003 & 0.1082 $\pm$ 0.0025 & 0.1000 $\pm$ 0.0018 & 0.0899 $\pm$ 0.0020 & 0.0953 $\pm$ 0.0027 \\
\OI$\lambda6300$ &-&-&-&-&-&-\\
\SIII$\lambda6312$ &-&-&-&-&-&-\\
$\mathrm{H_{\alpha}}$ & 2.79 $\pm$ 0.01 & 2.79 $\pm$ 0.02 & 2.79  $\pm$ 0.01 & 2.79 $\pm$ 0.01 & 2.79 $\pm$ 0.01  & 2.75 $\pm$ 0.02\\
\NII$\lambda6583$ & 0.0843 $\pm$ 0.0017 & 0.059 $\pm$ 0.003 & 0.0549 $\pm$ 0.0028 & 0.0831 $\pm$ 0.0021 & 0.1054 $\pm$ 0.0023 & 0.0888 $\pm$ 0.0027 \\
\SII$\lambda6716$ & 0.1840 $\pm$ 0.0017 & 0.1215 $\pm$ 0.0028 & 0.0899 $\pm$ 0.002 & 0.1764 $\pm$ 0.0018 & 0.2387 $\pm$ 0.0022 & 0.2229 $\pm$ 0.0026 \\
\SII$\lambda6731$ & 0.1281 $\pm$ 0.0013 & 0.0827 $\pm$ 0.0024 & 0.0626 $\pm$ 0.0015 & 0.1240 $\pm$ 0.0016 & 0.1675 $\pm$ 0.0017 & 0.1598 $\pm$ 0.0022 \\
\HeI$\lambda6678$ & 0.0264 $\pm$ 0.0014 & 0.028 $\pm$ 0.003 & 0.0277 $\pm$ 0.0024 & 0.0282 $\pm$ 0.0021 & 0.0185 $\pm$ 0.0018 & 0.0231 $\pm$ 0.0026 \\
\HeI$\lambda7065$ & 0.0190 $\pm$ 0.0016 & 0.024 $\pm$ 0.004 & 0.0206 $\pm$ 0.0026 & 0.0178 $\pm$ 0.0021 & 0.0127 $\pm$ 0.0017 & 0.0111 $\pm$ 0.0029 \\
\ArIII$\lambda7136$ & 0.0575 $\pm$ 0.0016 & 0.062 $\pm$ 0.005 & 0.0605 $\pm$ 0.0023 & 0.0361 $\pm$ 0.0019 & 0.0303 $\pm$ 0.0017 & 0.0281 $\pm$ 0.0021 \\
\hline
\end{tabular}
\end{table*}

\begin{table*}
\centering

\small
 \caption{\label{table_fluxes2}The same as Table \ref{tab:Te_Oabund_1}, but for regions 61, 63, 69, 1, 2.
\label{tab:Te_Oabund_2}}
\begin{tabular}{l|l|l|l|l|l}
\hline
region & 61 & 63 & 69 & 1 & 2 \\
\hline
\Ha/\Hb{} (obs. values)  & 3.05 $\pm$ 0.03 & 3.00  $\pm$ 0.01 & 2.84 $\pm$ 0.02 & 3.32 $\pm$ 0.05 & 3.16 $\pm$ 0.04  \\
\hline
\OII$\lambda3726,3729$ & 2.7 $\pm$ 0.3 & 1.77 $\pm$ 0.12 & 2.88 $\pm$ 0.18 & 2.8 $\pm$ 1.1 & 3.4 $\pm$ 0.9 \\
\NeIII$\lambda3869$ & 0.08 $\pm$ 0.07 & 0.262 $\pm$ 0.027 & 0.09 $\pm$ 0.03 &-&-\\
\HeI$\lambda3889$ & 0.13 $\pm$ 0.08 & 0.166 $\pm$ 0.020 & 0.11 $\pm$ 0.04 &-&-\\
$\mathrm{H_{\epsilon}}$ & 0.27 $\pm$ 0.07 & 0.322 $\pm$ 0.024 & 0.28 $\pm$ 0.03 &-&-\\
$\mathrm{H_{\delta}}$ & 0.220 $\pm$ 0.024 & 0.254 $\pm$ 0.012 & 0.241 $\pm$ 0.019 & 0.22 $\pm$ 0.06 & 0.18 $\pm$ 0.07 \\
$\mathrm{H_{\gamma}}$ & 0.451 $\pm$ 0.013 & 0.453 $\pm$ 0.007 & 0.433 $\pm$ 0.009 & 0.49 $\pm$ 0.04 & 0.496 $\pm$ 0.027 \\
\OIII$\lambda4363*$ & 0.038 $\pm$ 0.009 & 0.067 $\pm$ 0.003 & 0.035 $\pm$ 0.004 & 0.133 $\pm$ 0.021 & 0.066 $\pm$ 0.016 \\
\HeII &-&-&-& 0.054 $\pm$ 0.009 & - \\
\OIII$\lambda5007$ & 2.096 $\pm$ 0.018 & 3.134 $\pm$ 0.014 & 1.594 $\pm$ 0.008 & 3.37 $\pm$ 0.05 & 2.34 $\pm$ 0.03 \\
\HeI$\lambda5875$ & 0.098 $\pm$ 0.003 & 0.0951 $\pm$ 0.0016 & 0.1060 $\pm$ 0.0022 & 0.090 $\pm$ 0.006 & 0.098 $\pm$ 0.005 \\
\OI$\lambda6300$ & 0.0327 $\pm$ 0.0027 & 0.0577 $\pm$ 0.0014 & 0.0548 $\pm$ 0.0025 &-&-\\
\SIII$\lambda6312$ & 0.0119 $\pm$ 0.0015 & 0.0137 $\pm$ 0.0006 & 0.0076 $\pm$ 0.0010 &-&-\\
$\mathrm{H_{\alpha}}$ & 2.8 $\pm$ 0.02 & 2.78 $\pm$ 0.01 & 2.78 $\pm$ 0.01 & 2.74 $\pm$ 0.04 & 2.76 $\pm$ 0.03 \\
\NII$\lambda6583$ & 0.0743 $\pm$ 0.0025 & 0.0599 $\pm$ 0.0012 & 0.0941 $\pm$ 0.0021 & 0.104 $\pm$ 0.006 & 0.108 $\pm$ 0.005 \\
\SII$\lambda6716$ & 0.1856 $\pm$ 0.0029 & 0.1924 $\pm$ 0.0014 & 0.2494 $\pm$ 0.002 & 0.278 $\pm$ 0.006 & 0.232 $\pm$ 0.005 \\
\SII$\lambda6731$ & 0.1294 $\pm$ 0.0021 & 0.1362 $\pm$ 0.0012 & 0.1754 $\pm$ 0.0016 & 0.206 $\pm$ 0.005 & 0.157 $\pm$ 0.004 \\
\HeI$\lambda6678$ & 0.0278 $\pm$ 0.0029 & 0.0267 $\pm$ 0.0012 & 0.0366 $\pm$ 0.0021 & 0.037 $\pm$ 0.007 & 0.023 $\pm$ 0.005 \\
\HeI$\lambda7065$ & 0.0161 $\pm$ 0.0021 & 0.0178 $\pm$ 0.0011 & 0.0276 $\pm$ 0.0017 & 0.016 $\pm$ 0.003 & 0.011 $\pm$ 0.003 \\
\ArIII$\lambda7136$ & 0.0353 $\pm$ 0.0022 & 0.0399 $\pm$ 0.001 & 0.0270 $\pm$ 0.0016 & 0.033 $\pm$ 0.005 & 0.043 $\pm$ 0.004 \\
\hline
\end{tabular}
\end{table*}
\newpage
%%%%%%%%%%%%%%%%%%%%%%%%%%%%%%%%%%%%%%%%%%%%%%%%%%%%%%%%
%%%%%%%%%%%%%%%%%%%%%%%%%%%%%%%%%%%%%%%%%%%%%%%%%%%%%%%%

%%%%%%%%%%%%%%%%%%%%%%%%%%%%%%%%%%%%%%%%%%%%%%%%%%%%%%%%
%%%%%%%%%%%%%%%%%%%%%%%%%%%%%%%%%%%%%%%%%%%%%%%%%%%%%%%%
\begin{table*}

\small
\centering
\begin{minipage}{\linewidth}
% 
% \onecolumn
\caption{\label{table_fluxes_custom}The same as Table \ref{tab:Te_Oabund_1}. for regions 65-65C.
\label{tab:emission_lines_66_66C}}
\begin{tabular}{l|l|l|l|l}
\hline
region & 65 & 65A\footnote{We did not perform reddening correction for this region, but if correction is performed with E(B-V) of the nearby region 65, 12+log(O/H) for reg. 65A changes for no more than 7\%} & 65B & 65C \\
\hline
\Ha/\Hb{} (obs. values)  & 3.160 $\pm$ 0.007 & 2.385 $\pm$ 0.006 & 3.509 $\pm$ 0.008 &  3.096 $\pm$ 0.014  \\
\hline
\OII$\lambda3727$ & 0.417 $\pm$ 0.011 & 0.971 $\pm$ 0.026 & 0.441 $\pm$ 0.028 & 1.87 $\pm$ 0.16 \\
H12 & 0.014 $\pm$ 0.005 & 0.038 $\pm$ 0.013 & 0.008 $\pm$ 0.009 & - \\
H11 & 0.031 $\pm$ 0.003 & 0.070 $\pm$ 0.008 & 0.041 $\pm$ 0.009 & - \\
H10 & 0.0374 $\pm$ 0.0026 & 0.038 $\pm$ 0.007 & 0.025 $\pm$ 0.007 & - \\
H9 & 0.0598 $\pm$ 0.0019 & 0.086 $\pm$ 0.004 & 0.049 $\pm$ 0.004 & - \\
\NeIII$\lambda3869$ & 0.543 $\pm$ 0.004 & 0.712 $\pm$ 0.009 & 0.444 $\pm$ 0.008 & 0.37 $\pm$ 0.04 \\
\HeI$\lambda3889$ & 0.1936 $\pm$ 0.0026 & 0.283 $\pm$ 0.007 & 0.159 $\pm$ 0.006 & 0.21 $\pm$ 0.04 \\
$\mathrm{H_{\epsilon}}$ & 0.3081 $\pm$ 0.0028 & 0.407 $\pm$ 0.005 & 0.252 $\pm$ 0.005 & 0.281 $\pm$ 0.028 \\
\HeI$\lambda4026$ & 0.0161 $\pm$ 0.0012 & 0.0242 $\pm$ 0.0029 & 0.0110 $\pm$ 0.0028 & - \\
\SII$\lambda4069$ & 0.0052 $\pm$ 0.0008 & 0.007 $\pm$ 0.004 & 0.0058 $\pm$ 0.0022 & - \\
$\mathrm{H_{\delta}}$ & 0.2607 $\pm$ 0.0017 & 0.343 $\pm$ 0.004 & 0.228 $\pm$ 0.003 & 0.264 $\pm$ 0.022 \\
$\mathrm{H_{\gamma}}$ & 0.4898 $\pm$ 0.0017 & 0.5904 $\pm$ 0.0025 & 0.4359 $\pm$ 0.0021 & 0.489 $\pm$ 0.011 \\
\OIII$\lambda4363$ & 0.162 $\pm$ 0.001 & 0.1714 $\pm$ 0.0017 & 0.1315 $\pm$ 0.0014 & 0.085 $\pm$ 0.006 \\
\HeI$\lambda4387$ & 0.00495 $\pm$ 0.00025 & 0.0061 $\pm$ 0.0006 & 0.0045 $\pm$ 0.0006 & - \\
\HeI$\lambda4471$ & 0.0405 $\pm$ 0.0005 & 0.0473 $\pm$ 0.0011 & 0.033 $\pm$ 0.001 & 0.032 $\pm$ 0.004 \\
\FeIII$\lambda4658$ & 0.00146 $\pm$ 0.00015 & 0.0026 $\pm$ 0.0003 & 0.0021 $\pm$ 0.0003 & - \\
\HeII & 0.00417 $\pm$ 0.00013 & 0.0008 $\pm$ 0.0003 & 0.0123 $\pm$ 0.0003 & - \\
\ArIV$\lambda4711$ & 0.0309 $\pm$ 0.0003 & 0.0255 $\pm$ 0.0006 & 0.0242 $\pm$ 0.0005 & - \\
\ArIV$\lambda4740$ & 0.02043 $\pm$ 0.00023 & 0.0158 $\pm$ 0.0005 & 0.0146 $\pm$ 0.0004 & - \\
Hb & 1.0000 $\pm$ 0.0025 & 1.000 $\pm$ 0.003 & 1.0000 $\pm$ 0.0028 & 1.000 $\pm$ 0.006 \\
\OIII$\lambda5007$ & 7.126 $\pm$ 0.015 & 5.934 $\pm$ 0.015 & 6.815 $\pm$ 0.015 & 5.489 $\pm$ 0.025 \\
\HeI$\lambda5016$ & 0.0253 $\pm$ 0.0004 & 0.02292 $\pm$ 0.00015 & 0.0200 $\pm$ 0.0003 & 0.0233 $\pm$ 0.0006 \\
\ClIII$\lambda5518$ &0.0029 $\pm$ 0.00010 & 0.00346 $\pm$ 0.00016 & 0.00262 $\pm$ 0.00012 & - \\
\ClIII$\lambda5538$ & 0.0020 $\pm$ 0.00010 & 0.00248 $\pm$ 0.00015 & 0.00169 $\pm$ 0.00012 & - \\
\HeI$\lambda5876$ & 0.1144 $\pm$ 0.0003 & 0.1108 $\pm$ 0.0005 & 0.0971 $\pm$ 0.0004 & 0.1014 $\pm$ 0.0016 \\
\OI$\lambda6300$ & 0.0103 $\pm$ 0.0001 & 0.01540 $\pm$ 0.00028 & 0.00848 $\pm$ 0.00014 & 0.0289 $\pm$ 0.0025 \\
\SIII$\lambda6312$ & 0.01519 $\pm$ 0.00011 & 0.01847 $\pm$ 0.00027 & 0.01344 $\pm$ 0.00017 & 0.0202 $\pm$ 0.0015 \\
\OI$\lambda6364$ & 0.0035 $\pm$ 0.0001 & 0.00565 $\pm$ 0.00017 & 0.00293 $\pm$ 0.00011 & 0.0114 $\pm$ 0.0014 \\
$\mathrm{H_{\alpha}}$ & 2.780 $\pm$ 0.006 & 2.385 $\pm$ 0.006 & 2.790 $\pm$ 0.006 & 2.800 $\pm$ 0.013 \\
\NII$\lambda6584$ & 0.0184 $\pm$ 0.0002 & 0.0276 $\pm$ 0.0004 & 0.01817 $\pm$ 0.00027 & 0.0510 $\pm$ 0.0018 \\
\HeI$\lambda6678$ & 0.03165 $\pm$ 0.00015 & 0.0310 $\pm$ 0.0003 & 0.0270 $\pm$ 0.0002 & - \\
\SII$\lambda6716$ & 0.04139 $\pm$ 0.00016 & 0.0710 $\pm$ 0.0003 & 0.0377 $\pm$ 0.0002 & 0.1304 $\pm$ 0.0019 \\
\SII$\lambda6731$ & 0.03321 $\pm$ 0.00012 & 0.05250 $\pm$ 0.00027 & 0.02865 $\pm$ 0.00016 & 0.0927 $\pm$ 0.0015 \\
\HeI$\lambda7065$ & 0.03369 $\pm$ 0.00026 & 0.0273 $\pm$ 0.0003 & 0.02473 $\pm$ 0.00019 & 0.0178 $\pm$ 0.0012 \\
\ArIII$\lambda7136$ & 0.0503 $\pm$ 0.0007 & 0.0586 $\pm$ 0.0006 & 0.0423 $\pm$ 0.0004 & 0.0537 $\pm$ 0.0018 \\
\hline
\end{tabular}
\end{minipage}
\end{table*}
%%%%%%%%%%%%%%%%%%%%%%%%%%%%%%%%%%%%%%%%%%%%%%%%%%%%%%%%
%%%%%%%%%%%%%%%%%%%%%%%%%%%%%%%%%%%%%%%%%%%%%%%%%%%%%%%%
\newpage

\begin{table*}
\caption{Parameters of best-fit components for \OIII{} and \Ha{} lines for regions 65-65C (corresponding to Mrk~71), where $\sigma$ is velocity dispersion in $\rm km/s$,  $v$ is the deviation from the main component of the emission line ($km/s$), and A is intensity of the line in $10^{-15}~\rm erg/cm^2/s/$\AA. The main component of \Ha{} has complex structure  due to instrumental effects and therefore fitted by two Gaussians. The corresponding fit for region 65 could be found in Fig.\ref{fig:66}.\label{tab:kin_par}}
\begin{tabular}{l|l|l|l|l|l}
line                               & parameter & 65   & 65A  & 65B  & 65C   \\
\hline
\multirow{3}{*}{$\mathrm{[OIII]_1}$(main) }& $\sigma$         & 111  & 111  & 111  & 112   \\
                                   & $v$        & -    & -    & -    & -     \\
                                   & A         & 102  & 48   & 71   & 5.3  \\
                                   \hline
\multirow{3}{*}{$\mathrm{[OIII]_2}$}           & $\sigma$         & 540  & 539  & 539  & -     \\
                                   & $v$       & -4   & 3    & -5   & -     \\
                                   & A         & 0.25 & 0.20 & 0.25 & -     \\
                                   \hline
\multirow{3}{*}{$\mathrm{[OIII]_3}$}           & $\sigma$         & 1918 & 1918 & 1918 & 2097  \\
                                   & $v$        & -64  & 3    & -65  & -70   \\
                                   & A         & 0.06 & 0.12 & 0.14 & 0.04  \\
                                   \hline
\multirow{3}{*}{$\mathrm{H_{\alpha1.1}}$(main) }             & $\sigma$         & 88   & 86   & 88   & 88    \\
                                   & $v$        & 0    & 0    & 0    & 0     \\
                                   & A         & 42   & 17   & 33   & 2.7  \\
                                   \hline
\multirow{3}{*}{$\mathrm{H_{\alpha1.2}}$(main)}             & $\sigma$         & 137  & 137  & 137  & 137   \\
                                   & $v$        & 229  & 175  & 174  & 192   \\
                                   & A         & 1    & 1    & 1    & 0.07  \\
                                   \hline
\multirow{3}{*}{$\mathrm{H_{\alpha2}}$}             & $\sigma$        & 640  & 823  & 823  & 823   \\
                                   & $v$       & 105  & 175  & 60   & 55    \\
                                   & A         & 0.10 & 0.08 & 0.10 & 0.015
\end{tabular}
\end{table*}

% Don't change these lines
\bsp	% typesetting comment
\label{lastpage}
\end{document}